\documentclass{desyproc}
\usepackage{color}
\usepackage{amssymb}
\usepackage{amsmath}
\usepackage{epsfig}
\usepackage{subfigure}
\usepackage{mathrsfs}
\usepackage{longtable}
\usepackage[usenames,dvipsnames]{xcolor}
\usepackage[linktoc=all]{hyperref}
\usepackage{titlesec}
\usepackage{lscape}
\usepackage[official]{eurosym}
\usepackage[utf8]{inputenc}

\newcommand{\be}{\begin{equation}}
\newcommand{\ee}{\end{equation}}

\begin{document}
%------------------------------------
\title{Axions and ALPs: a very short introduction}

%for single authors the superscripts are optional
\author{{\slshape David J. E. Marsh${}^{1,2}$}\\[1ex]
$^1$Department of Physics, King's College London, United Kingdom. \\
$^2$Institut f\"{u}r Astrophysik, Georg-August Universitat, G\"{o}ttingen, Germany.}

% if the proceedings are available online (e.g. at Indico)
% please enter the contribution ID or file_name below for the DOI
%\contribID{32}

% TO THE CONFERENCE EDITORS: 
% please update the following information      
% before sending the template to the authors
%\confID{16884}  % if the conference is on Indico uncomment this line
%\desyproc{DESY-PROC-2017-02}
\acronym{Patras 2017} % if you want the Acronym in the page footer uncomment this line
%\doi  % if there is an online version we will register DOIs

\maketitle

\begin{abstract}
The QCD axion was originally predicted as a dynamical solution to the strong CP problem. Axion like particles (ALPs) are also a generic prediction of many high energy physics models including string theory. Theoretical models for axions are reviewed, giving a generic multi-axion action with couplings to the standard model. The couplings and masses of these axions can span many orders of magnitude, and cosmology leads us to consider several distinct populations of axions behaving as coherent condensates, or relativistic particles. Light, stable axions are a mainstay dark matter candidate. Axion cosmology and calculation of the relic density are reviewed. A very brief survey is given of the phenomenology of axions arising from their direct couplings to the standard model, and their distinctive gravitational interactions.\end{abstract}

\section{Theory of Axions}

\subsection{The QCD Axion}

The QCD axion was introduced by Peccei \& Quinn~\cite{pecceiquinn1977}, Weinberg~\cite{weinberg1978}, and Wilczek~\cite{wilczek1978} (PQWW) in 1977-78 as a solution to the CP problem of the strong interaction. This arises from the Chern-Simons term:
\be
\mathcal{L}_{\theta{\rm QCD}}=\frac{\theta_{\rm QCD}}{32\pi^2}\text{Tr } G_{\mu\nu}\tilde{G}^{\mu\nu}\, ,
\ee
where $G$ is the gluon field strength tensor, $\tilde{G}^{\mu\nu}=\epsilon^{\alpha\beta\mu\nu}G_{\alpha\beta}$ is the dual, and the trace runs over the colour $SU(3)$ indices. This term is called topological since it is a total derivative and does not affect the classical equations of motion. However, it has important effects on the quantum theory. This term is odd under CP, and so produces CP-violating interactions, such as a neutron electric dipole moment (EDM), $d_n$. The value of $d_n$ produced by this term was computed in Ref.~\cite{1979PhLB...88..123C} to be 
\be
d_n\approx 3.6 \times 10^{-16}\theta_{\rm QCD} \, e\text{ cm}\, ,
\ee
where $e$ is the charge on the electron. The (permanent, static) dipole moment is constrained to $|d_n|<3.0\times 10^{-26}\,e\,\text{cm}$ (90\% C.L.)~\cite{2015PhRvD..92i2003P}, implying $\theta_{\rm QCD}\lesssim 10^{-10}$. 

If there were only the CP-conserving strong interactions, then $\theta_{\rm QCD}$ could simply be set to zero by symmetry. In the real world, and very importantly, the weak interactions violate CP~\cite{2014ChPhC..38i0001O}. By chiral rotations of the quark fields, we see that the physically measurable parameter is
\be
\theta_{\rm QCD}=\tilde{\theta}_{\rm QCD}+\text{arg det}M_uM_d\, ,
\ee
where $\tilde{\theta}$ is the ``bare'' (i.e. pure QCD) quantity and $M_u$, $M_d$ are the quark mass matrices. Thus the smallness of $\theta_{\rm QCD}$ implied by the EDM constraint is a fine tuning problem since it involves a precise cancellation between two dimensionless terms generated by different physics.

The famous PQ solution to this relies on two ingredients: the Goldstone theorem, and the presence of \emph{instantons} in the QCD vacuum. A global chiral $U(1)_{\rm PQ}$ symmetry, is introduced, under which some quarks are charged. This symmetry is spontaneously broken by a scalar field. More precisely, the symmetry breaking potential for the complex scalar $\varphi$ is:
\be
V(\varphi)=\lambda(|\varphi|^2-f_a^2/2)^2\, \Rightarrow \langle\varphi\rangle=(f_a/\sqrt{2})e^{i \phi/f_a} .
\ee
The Goldstone boson is $\phi$, the axion, and I have defined the vacuum expectation value (VEV) of the field to be $f_a/\sqrt{2}$ to give canonical kinetic terms. $f_a$ is known as the axion ``decay constant'' (we will shortly see the analogy to pions). 

The charges, $\mathcal{Q}_{\rm PQ}$ of some quarks (either the standard model quarks or new heavy objects with colour charge) under $U(1)_{\rm PQ}$ are such that the PQ symmetry is anomalous, with the colour anomaly given by~\cite{Srednicki:1985xd}:
\be
\mathcal{C}\delta_{ab}=2\text{Tr }\mathcal{Q}_{\rm PQ}T_aT_b \, .
\ee
The trace is over all the quarks, and the $T_a$ are the generators for the representations of the quarks under $SU(3)$. An anomalous chiral rotation by $\phi/f_a$ of the quarks changes the fermion measure in the path integral, and leads to a change in the action:
\be
S\rightarrow S+\int d^4x\frac{\mathcal{C}}{32\pi^2}\frac{\phi}{f_a}{\rm Tr}\,G_{\mu\nu}\tilde{G}^{\mu\nu}\, .
\ee
For the QCD axion it is common to absorb the colour anomaly into the definition of $f_a$, and keep the VEV a separate quantity $v_{\rm PQ}$. However, I find it more useful, especially when considering multi-axion theories, to keep the anomaly factors explicit.

\begin{figure}
\begin{center}
\includegraphics[scale=0.5]{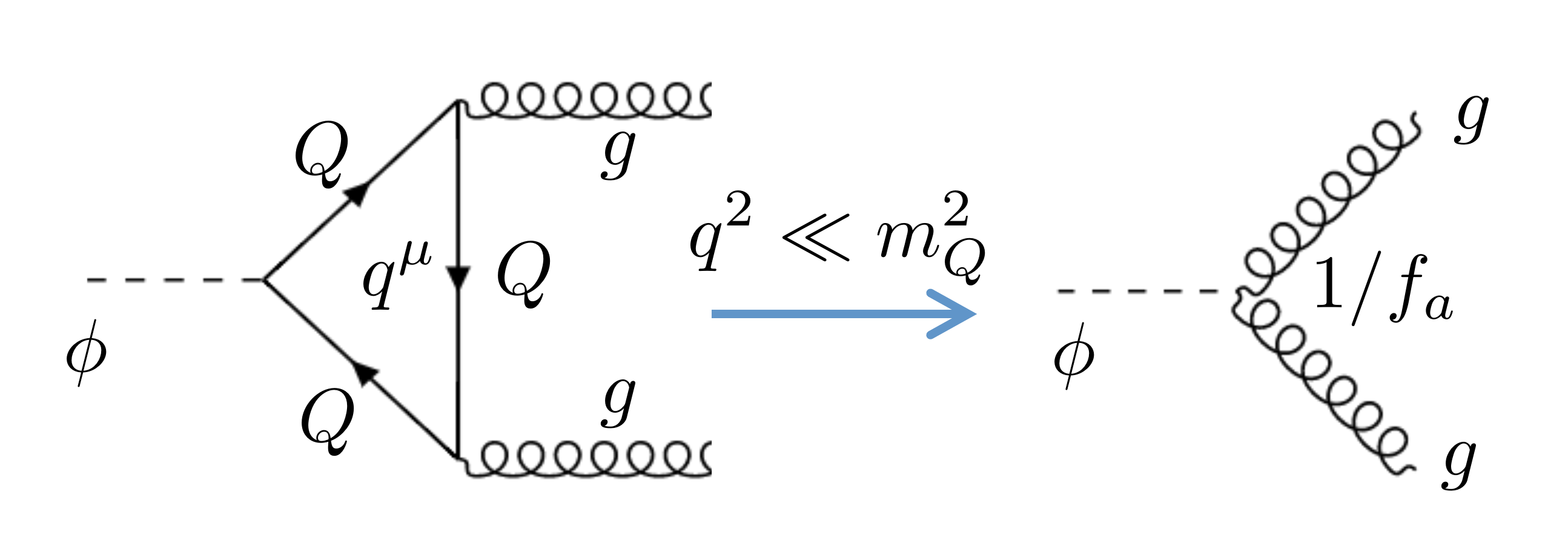}
 \end{center}
 \caption{The colour anomaly in the KSVZ axion model. Heavy quarks, $Q$, run in a loop with momentum $q^\mu$. At low-momentum transfer, $q^2\ll m_Q^2$, the interaction can be replaced with the effective $\phi G\tilde{G}/f_a$ interaction. Reproduced from Ref.~\cite{2016PhR...643....1M}.}
\label{fig:ksvz_triangle}
\end{figure}

The colour anomaly can be understood via the Feynman diagram in Fig.~\ref{fig:ksvz_triangle}: the loop of quarks, which are \emph{chirally} charged under $U(1)_{\rm PQ}$, mediate an interaction between the axion and the gluons once PQ symmetry is broken. An accessible description of the computation of the axial anomaly can be found in  Zee's book~\cite{2003qftn.book.....Z}. Anomalies arise when the quantum theory does not obey a symmetry of the classical theory. In this case the symmetry is the conservation of the axion current, which is violated by the production of two gluons. Anomalies can be understood via the path integral as due to the change in the fermion measure in the partition function.

Now, via the anomaly, the axion is coupled to the QCD Chern-Simons term. Since the only other term in the axion action is the kinetic term, we are free to shift the axion field by an arbitrary constant and absorb the value of $\theta_{\rm QCD}$ into $\phi$ by a field redefinition. The physical observable related to CP violation in the strong interactions is given by $\phi/f_a$, which is \emph{dynamical}. 

As promised, the final ingredient of the PQ theory comes from instantons. Instantons are solutions to the Euclidean equations of motion that cause the vacuum energy to depend on $\theta=\phi/f_a$. In the dilute instanton gas (DIG, e.g. Ref.~\cite{1988assy.book.....C}) approximation the dependence is $E_{\rm vac}\propto (1-\cos \mathcal{C}\theta)$. This energy dependence means that we can write a quantum effective action that includes a potential for the axion. This potential is, happily, minimized at the CP-conserving value $\mathcal{C}\theta=0$ (mod $2\pi$), and so the axion \emph{dynamically} solves the problem of the smallness of $\theta$. The proof for the CP conservation of the instanton corrected action is known as the Vafa-Witten theorem~\cite{Vafa:1984xg}.

The axion mass induced by the interaction with QCD was famously computed by Weinberg~\cite{weinberg1978} and Wilczek~\cite{wilczek1978} using chiral perturbation theory (ChPT):
\be
m_{a,{\rm QCD}}\approx 6\times 10^{-6}\text{ eV}\frac{10^{12}\text{ GeV}}{f_a/\mathcal{C}} \, ,
\ee 
with the full ChPT potential at zero temperature given by (e.g. Ref.~\cite{2016JHEP...01..034D})
\be
V(\phi)=-m_\pi^2 f_\pi^2\sqrt{1-4z \sin^2(\mathcal{C}\phi/2f_a)}\,;\quad z=\frac{m_um_d}{(m_u+m_d)^2}\, .
\ee
Note that this potential differs from the DIG result, $V(\phi)\propto \cos (\phi/f_a)$, and that it vanishes in the limit of massless quarks.\footnote{Hence, if it were experimentally consistent to have a massless up or down quark, then there would be no strong CP problem. Chiral rotation of the massless quark could absorb the problematic term.} 

The temperature dependence of the axion potential is expressed through the topological susceptibilty of QCD, $\chi(T)$.  The axion potential is $V(\phi)=\chi(T)U(\theta)$ where $T$ is temperature and $U(\theta)$ is a dimensionless periodic function. Using that $U(\theta)$ is quadratic about the minimum for a massive particle we see that $\chi(T)=m_a^2(T)f_a^2$, and so we often talk instead of the temperature dependence of the axion mass. It is common to parameterise the dependence by a (possibly varying) power law: :
\be
m_a(T)= m_{a,0} \left(\frac{T}{\Lambda_{\rm QCD}}\right)^{-n} \, ; \quad (T\gg \Lambda_{\rm QCD}) \, ,
\ee
with the mass approaching the zero temperature value for $T<\Lambda_{\rm QCD}$. At lowest order the DIG gives the famous result $n=4$ for QCD in the standard model~\cite{1981RvMP...53...43G}.\footnote{For a general Yang-Mills theory of gauge group $SU(N_c)$ with $N_f$ quark flavours the index is $n= (11 N_c-2N_f)/6+N_f/2-2$. The number of flavours is the number of active flavours, i.e. those lighter than the confinement scale, which for QCD in the standard model is $N_f=3$ for up, down, strange.} In the recent lattice QCD calculations of Refs.~\cite{2016PhLB..752..175B,Borsanyi:2016ksw} the index is consistent with the DIG at high temperature, while the low temperature behaviour (relevant for predicting the onset of axion oscillations) is better fit by $n=3.55\pm 0.3$.

Because it descends as the argument of the complex field $\varphi$, the values $\phi$ and $\phi+2\pi f_a$ are physically equivalent (in the absence of monodromy in the complex plane). However the effective potential $V(\phi)$ has minima at CP conserving values $\phi+2\pi f_a/\mathcal{C}$. This implies that the PQ charges must be normalised such that $\mathcal{C}$ is an integer~\cite{Srednicki:1985xd}. Thus there are $\mathcal{C}$ distinct vacua, which lead to $\mathcal{C}$ distinct types of domain wall solution~\cite{Sikivie:1982qv}. It is therefore common to denote $\mathcal{C}=N_{\rm DW}$ as the domain wall number.

The PQ symmetry is also anomalous with respect to $U(1)_{\rm EM}$, with the electromagnetic anomaly given by 
\be
\mathcal{E}=2\text{Tr } \mathcal{Q}_{\rm PQ}\mathcal{Q}_{\rm EM}^2\, ,
\ee
and $\mathcal{Q}_{\rm EM}$ are the EM charges of the fermions. This anomaly introduces a coupling to electromagnetism:
\be
\mathcal{L}_{\rm int}\supset-\frac{g_{\phi\gamma}}{4}\phi F_{\mu\nu}\tilde{F}^{\mu\nu}\, ,
\ee
with (e.g. Ref.~\cite{Srednicki:1985xd})
\be
g_{\phi\gamma}=\frac{\alpha_{\rm EM}}{2\pi f_a}\left( \mathcal{E} -\mathcal{C}\frac{2}{3}\cdot\frac{4+m_u/m_d}{1+m_u/m_d}\right)\, .
\label{eqn:photon_coupling}
\ee
The second half of this interaction arises after chiral symmetry breaking due to the colour anomaly and mixing with the $Z$, and is the preserve of the QCD axion. The first half of the interaction is allowed for any ALP.

The EM interaction mediates axion decay to two photons with lifetime:
\be
\tau_{\phi\gamma}=\frac{64\pi}{m_a^3g_{\phi\gamma}^2} \approx 130\text{ s}\left(\frac{\text{GeV}}{m_a}\right)^3\left(\frac{10^{-12}\text{ GeV}^{-1}}{g_{\phi\gamma}}\right)^2\, ,
\label{eqn:decay_lifetime}
\ee
hence why $f_a$ is referred to as the decay constant. This interaction historically rules out the original PQWW axion, where $f_a$ is tied to the weak scale, from e.g. beam dump experiments~(see e.g. Refs.~\cite{Kim:1986ax,2015PhRvD..92b3010M}). 

Viable QCD axion models are split into two canonical types: ``KSVZ''~\cite{1979PhRvL..43..103K,1980NuPhB.166..493S}, which mediate the anomaly through additional heavy quarks, and ``DFSZ''~\cite{1981PhLB..104..199D,Zhitnitsky:1980tq} which mediate the anomaly through the standard model quarks. There are, however, a large number of possible variations on these themes, which allow a wide range of possible couplings between the axion and the standard model~\cite{2017PhRvL.118c1801D}, even in the restricted class of a single axion with mass arising from QCD instantons alone. Theories of multiple ALPs, to which we now turn in a string theory context, allow for even more variation.

\subsection{Axions in Supergravity and String Theory}

This section is intended only to give a flavour for what is, unsurprisingly, a very complicated story. For more details, see Refs.~\cite{2006JHEP...06..051S,2006JHEP...05..078C,2007stmt.book.....B,2012arXiv1209.2299R,2012JHEP...10..146C}. Consider the ten dimensional effective supergravity action for a $p$-form field $A_{p}$ with field strength $F_{p+1}={\rm d}A_p$:\footnote{Differential form notation for the uninitiated physicist is introduced in Refs.~\cite{2004sgig.book.....C,1987cup..bookR....G}.}
\be
S_{\rm 10D}=-\frac{1}{2}\int F_{p+1}\wedge \star F_{p+1}\, .
\ee
We dimensionally reduce this action on a 6-manifold $X$ by writing the field $A_{p}$ as a sum of \emph{harmonic} $p$-forms on $X$, which form a complete basis:
\be
A_p=\sum_{i}^{b_p}a_i(x)\omega_{p,i}(y)\, .
\label{eqn:harmonic_basis}
\ee 
The co-orindates $x$ are in the large $3+1$ dimensions, while $y$ are in the compact dimensions of $X$. Since $\omega$ is harmonic, ${\rm d}\omega=0$, the equations of motion on $X$ are automatically satisfied. The fields $a_i$ are the axion fields, which appear as pseudo-scalars in the dimensionally reduced action, with a shift symmetry descending from the gauge invariance of $F_{p+1}$ in ten dimensions. The axions are related to the integrals of the $p$-form as:
\be
a_i = \int _{C_{p,i}}A_p \, ,
\ee
where $C_{p,i}$ is the $i$th closed $p$-cycle on $X$. At this stage the axions are dimensionless angular variables and are not canonically normalised. The normalisations are fixed by the moduli of $X$.

The sum in Eq.~\eqref{eqn:harmonic_basis} extends over the number of harmonic $p$-forms on $X$, which is determined by the topology and expressed as the $p$th Betti number, $b_p$. For the standard phenomenology of string theory, with $\mathcal{N}=1$ supersymmetry in 3+1 dimensions the manifold $X$ must be so-called \emph{Calabi-Yau}~\cite{1985NuPhB.258...46C}, and the Betti numbers are given by the Hodge numbers $h_{1,1}$ and $h_{1,2}$. The properties of such manifolds have been studied in great detail~\cite{Altman:2014bfa,He:2017aed}. 
\begin{figure}
\begin{center}
\includegraphics[scale=0.45]{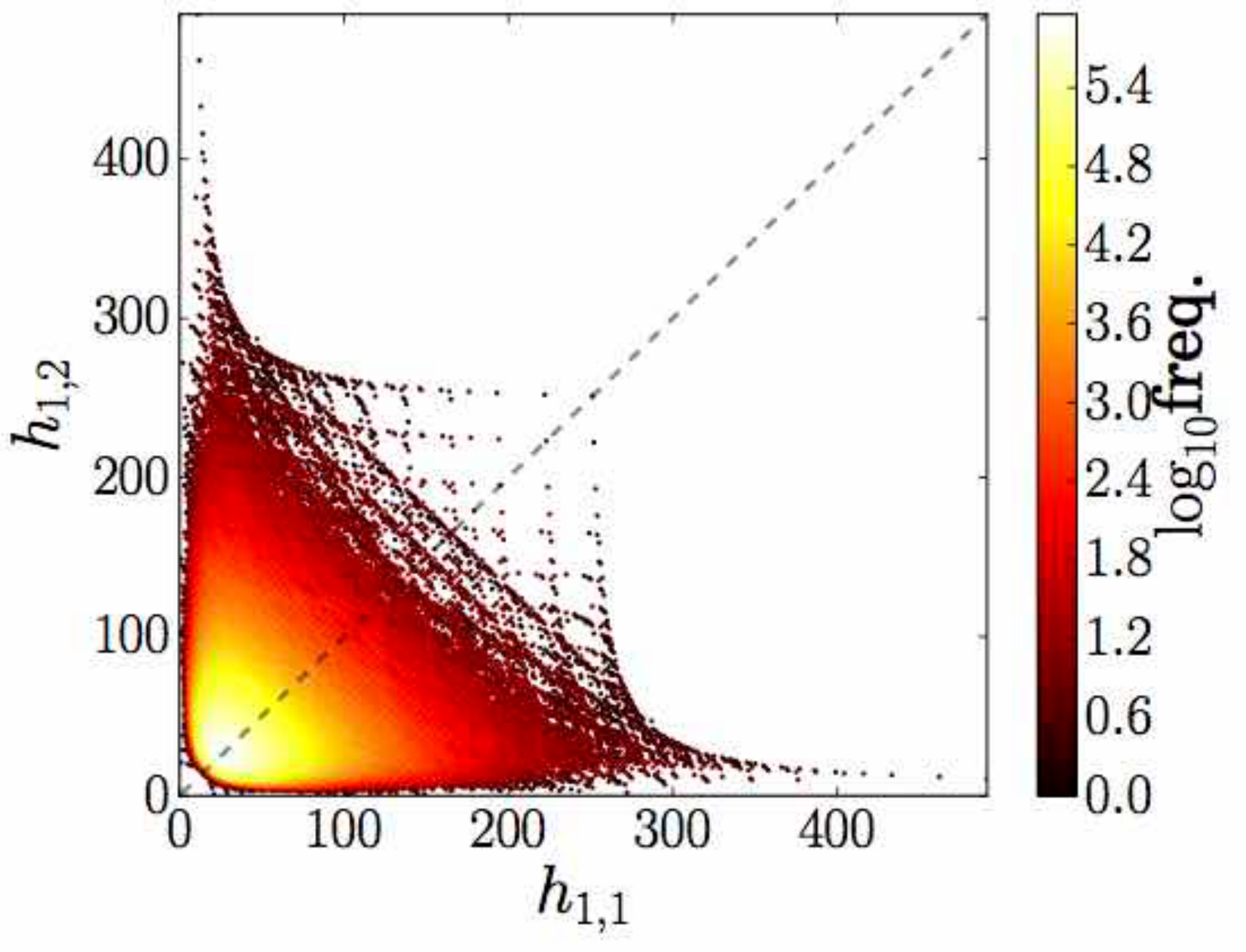}
\caption{The distribution of Hodge numbers $h_{1,1}$ and $h_{1,2}$ for the Calabi-Yau manifolds in the Kreuzer-Skarke~\cite{Kreuzer:2000xy} list. The peak in the distribution implies that a ``random'' Calabi-Yau sting vacuum will contain of order 30 axions. Reproduced from Ref.~\cite{2016PhR...643....1M}.}
\label{fig:kreuzer_skarke}
\end{center}
\end{figure}

For example, the Hodge numbers of 473,800,776 Calabi-Yau manifolds are known from the construction of all reflexive polyhedra in four dimensions performed by Kreuzer and Skarke~\cite{Kreuzer:2000xy,2000math......1106K}. The distribution in the plane $(h_{1,1},h_{1,2})$ of such manifolds is shown in Fig.~\ref{fig:kreuzer_skarke} and displays remarkable symmetry. The two most striking features are the symmetry about the axis $h_{1,1}=h_{1,2}$, known as mirror symmetry, and the large peak in the distribution near $h_{1,1}=h_{1,2}\approx 30$. The huge peak in the distribution implies that \emph{a random Calabi-Yau manifold constructed in this way is overwhelmingly likely to contain of order 30 axions}. This is the origin of the common lore that ``string theory predicts a large number of ALPs''.

To progress further, we must see if string theory tells us anything about the values of the axion masses or decay constants. We can assess the approximate scaling of these quantities in the example of Type-IIB theory compactified on orientifolds~\cite{Grimm:2004uq}.\footnote{I will assume throughout that the moduli have been stabilised with masses larger than the axions. There are important subtleties related to the scheme for moduli stabilisation and supersymmetry breaking, which alter the number of axions in the low energy theory. I will ignore these subtleties for simplicity of presentation, but the following cannot be considered a complete model in any sense.} The four-dimensional effective axion action coming from the $C_4$ field contains the kinetic term: 
\be
S_{4{\rm D}}=-\frac{1}{8}\int {\rm d}a_i\mathcal{K}_{ij}\wedge\star{\rm d}a_j \, ; \quad \mathcal{K}_{ij}=\frac{\partial^2K}{\partial \sigma_i\partial \sigma_j} \, ,
\ee
where $\sigma_i$ are the moduli, and $K$ is the K\"{a}hler potential. By canonically normalising the $a$ fields as $f_a^2(\partial a)^2$ we see that the decay constants are the eigenvalues of the K\"{a}hler metric and they scale like $f_a\sim M_{pl}/\sigma$. For the supergravity approximation to hold we must be at $\sigma>1$ in string units, and so \emph{the decay constants are parametrically sub-Planckian} (for $\sigma<1$ there is a $T$-dual description). 

As an example, consider axion masses arising from instantons of a non-Abelian gauge group (just like in QCD). Such a group can be realised by wrapping D7 branes on 4-cycles in $X$ (the same 4-cycles that we compactified $C_4$ on to obtain the axions) with the remaining part of the world-volume filling the non-compact dimensions. The super potential induced by the instantons is~\cite{2006JHEP...06..051S}:
\be
W=-M^3e^{-S_{\rm inst}+ia} \Rightarrow V(\phi)=-m_{\rm SUSY}^2M_{pl}^2e^{-S_{\rm inst}}\cos (\phi/f_a) \, ,
\ee
where $M$ is the scale of the instanton physics. The axion mass is exponentially sensitive to the instanton action, and scales as:
\be
m_a\sim m_{\rm SUSY} \frac{M_{pl}}{f_a} e^{-S_{\rm inst}/2}\, ,
\ee
with $m_{\rm SUSY}$ the scale of supersymmetry breaking. ``Typical'' instanton actions $S_{\rm inst.}\sim\mathcal{O}(100)$~\cite{2006JHEP...06..051S} lead to \emph{parametrically light axions}. The instanton action itself scales with the gauge coupling of the group, which is determined by the moduli and scales as:
\be
S_{\rm inst}\sim \frac{1}{g^2}\sim \sigma^2 \, .
\ee
Thus, as the different moduli take different values, so \emph{the axion masses can span many orders of magnitude}.

\subsection{The Multi Axion Effective Action}

The paramteric scalings above are a useful guide to think about axions in string theory, and are the essential basis for the popular phenomenology of the ``string axiverse''~\cite{axiverse}. However, a theory of multiple ALPs, whether it be inspired by string theory or not, must account for the fact that both the kinetic matrix (which may or may not be the K\"{a}hler metric) and the mass matrix are indeed matrices. Thus, the distributions of axion masses and decay constants are determined not by simple scalings for a single particle, but by the properties of the eigenvalues of (possibly large, possibly random) matrices~\cite{2006JCAP...05..018E,2017PhRvD..96h3510S}.

The general action before chiral symmetry breaking, but below all PQ scales, moduli masses, and the compactification scale is:
\begin{align}
\mathcal{L}=&-K_{ij}\partial_\mu\theta_i\partial^\mu\theta_j - \sum_{n=1}^{N_{\rm inst.}-1}\Lambda_n U_n( \mathcal{Q}_{i,n}\theta_i+\delta_n) \nonumber\\
 &-\frac{1}{4}\tilde{c}^{\rm EM}_i\theta_iF_{\mu\nu}\tilde{F}^{\mu\nu}-\frac{1}{4}c^{\rm QCD}_i\theta_iG_{\mu\nu}\tilde{G}^{\mu\nu} \nonumber\\
& +c^{q}_i\partial_\mu\theta_i(\bar{q}\gamma^\mu\gamma_5q)+c^{e}_i\partial_\mu\theta_i(\bar{e}\gamma^\mu\gamma_5e)\, .
\end{align}
The sums in $i$ and $j$ implied by repeated indices extend from $1$ to $N_{\rm ax}$, the number of light axions (with ``light'' defined, of course, by the scale of the problem, which could range from the Planck scale to the Hubble scale).

The first term is the general kinetic term which includes mixing of different axions. In this notation $K_{ij}$ has mass dimension two and contains off diagonal terms. 

The second term is the most general instanton potential, with $U_n$ an arbitrary periodic function. The sum extends over the number of instantons. The matrix $\mathcal{Q}$ is the instanton charge matrix (see e.g. Ref.~\cite{Bachlechner:2017zpb}). For gauge theory instantons, the entries of $\mathcal{Q}$ are determined by the chiral anomaly of the gauge group under each PQ symmetry. 

Since before chiral symmetry breaking I have included the axion-gluon coupling, the sum over instantons at first excludes the QCD contribution to the axion potential (though other instantons may also have temperature dependence that switch on only at lower temperatures, a [pssibility we ignore here). For \emph{any} theory of quantum gravity, there always exists the so-called ``axion wormhole'' instanton~\cite{1988NuPhB.306..890G,2017arXiv170607415A} and thus $N_{\rm inst}\geq N_{\rm ax}$. We allow arbitrary phases for each instanton, some of which can be absorbed by shifts in the axions, leaving $N_{\rm inst}-N_{\rm ax}\geq 0$ physical phases.

The next terms are the couplings between the axions and the standard model. I have considered only the coupling of axions to the light degrees of freedom of the standard model excluding neutrinos, since these are the ones relevant for experiment. 

Next, chiral symmetry breaking occurs, and the action changes:
\begin{align}
\mathcal{L}=&-K_{ij}\partial_\mu\theta_i\partial^\mu\theta_j - \sum_{n=1}^{N_{\rm inst.}}\Lambda_n U_n( \mathcal{Q}_{i,n}\theta_i+\delta_n) \nonumber\\
 &-\frac{1}{4}c^{\rm EM}_i\theta_iF_{\mu\nu}\tilde{F}^{\mu\nu}-\frac{i}{2}c^{d}_i\theta_i\bar{N}\sigma_{\mu\nu}\gamma_5NF^{\mu\nu} \nonumber\\
& +c^{N}_i\partial_\mu\theta_i(\bar{N}\gamma^\mu\gamma_5N)+c^{e}_i\partial_\mu\theta_i(\bar{e}\gamma^\mu\gamma_5e)
\label{eqn:largangain_post_chsb}
\end{align}

The potential has now been shifted by the QCD instanton contribution. In general there may be more than one axion with a non-zero colour anomaly. The electromagnetic coupling of the axion in the third term is shifted by the colour anomaly contribution, as in Eq.~\eqref{eqn:photon_coupling}. The fourth term is the induced coupling between the axions and the nucleon EDMs proportional to $\mathcal{C}_i$, and the fifth term is the coupling to the nucleon axial current. For more detail on the couplings, see Refs.~\cite{Srednicki:1985xd,2013PhRvD..88c5023G}.

Diagonalising the kinetic term first by the rotation matrix $U$, we see that the decay constants are given by the eigenvalues: $\vec{f}_{a}=\sqrt{2{\rm eig}(K)}$. The masses are found by diagonalising the matrix $\tilde{M}=2\text{diag}(1/f_a)UMU^T\text{diag}(1/f_a)$ with a rotation $V$, where $M$ is the mass matrix of Eq.~\eqref{eqn:largangain_post_chsb}. The canonically normalised field $\vec{\phi}=M_{pl}V\text{diag}(f_a)U\vec{\theta}$ has Lagrangian:
\begin{align}
\mathcal{L}=\sum_{i=1}^{N_{\rm ax}} & [-\frac{1}{2}\partial_\mu\phi_i\partial^\mu\phi_i-m_i^2\phi_i^2 \nonumber\\
 &-\frac{g_{i,\gamma}}{4}\phi_iF_{\mu\nu}\tilde{F}^{\mu\nu}-\frac{i}{2}g_{i,d}\phi_i\bar{N}\sigma_{\mu\nu}\gamma_5NF^{\mu\nu} \nonumber\\
& +\frac{g_{i,N}}{2m_N}\partial_\mu\phi_i(\bar{N}\gamma^\mu\gamma_5N)+\frac{g_{i,e}}{2m_e}\partial_\mu\phi_i(\bar{e}\gamma^\mu\gamma_5e) \, ]-V_{\rm int.}(\vec{\phi}). 
\end{align}

To assess whether this theory still solves the strong CP problem, we must consider the linear combination of axions that couples to the neutron EDM, its effective potential, and its VEV. Additional instantons, and other contributions to the potential, can spoil the solution by shifting the minimum. In the cosmological evolution of the axion field the temperature dependence of each term must also be considered.

The instanton-inspired form for the potential applies for ``true'' axions which obey a discrete shift symmetry. For so-called ``accidental axions''~\cite{2014JHEP...06..037D} additional contributions to the potential arise at the scale where the shift symmetry is explicitly broken, for example if the true symmetry is a global discrete symmetry like $\mathbf{Z}_N$~\cite{2016PhRvD..93b5027K}. If the shift symmetry undergoes a monodromy, then further explicit breaking can be induced~\cite{2008PhRvD..78j6003S,2010PhRvD..82d6003M}. Often such a spoiling of the axion symmetry is thought of in terms of the contribution of Planck suppressed operators to the action, under the common lore that ``quantum gravity violates all continuous global symmetries''~\cite{1992PhLB..282..137K}. Understanding the axion wormhole instanton leads to a more subtle view of this point, since the symmetry breaking is in fact non-perurbative~\cite{1995PhRvD..52..912K,2017arXiv170607415A}.

%%%%%%%%%%%%%%%%%%%%%%%%%%%%%
\section{Axion Cosmology}

\subsection{Axion Populations}

In the following we consider only axions that are stable on a Hubble time. There are four sources of cosmic axion energy density:
\begin{itemize}
\item Coherent displacement of the axion field. This accounts for the so-called misalignment populations of dark matter axions, and also for axion quintessence and axion inflation.
\item Axions produced via the decay of a topological defect. The topological defect is a configuration of the PQ field. When the defect decays, it produces axions.
\item Decay of a parent particle. Heavy particles such as moduli can decay directly into axions. If the mass of the parent is much larger than the axion, then the produced particles are relativistic.
\item Thermal production. Axions are coupled to the standard model. If the couplings are large enough, a sizeable population of thermal axions is produced.
\end{itemize} 
While the first two populations are sometimes thought of as distinct, in fact they are not. In a complete classical simulation of the defects directly from the PQ field, axion production is captured by the coherent field oscillations set up when the defect becomes unstable. The reason for the separation is that defects such as strings are sometimes more easily simulated using an effective description such as the Nambu-Goto action, in which case string decay and axion production must be added in as an additional effect.

These different axion populations manifest different phenomenology in cosmology:
\begin{itemize}
\item Coherent effects. The axion field only behaves as cold, collisionless particles on scales larger than the coherence length. This leads to wavelike effects on scales of order the de Broglie wavelength, and ``axion star'' formation that both distinguish axions from weakly interacting massive particles. These effects are particularly pronounced when the axion mass is very small, $m_a\sim 10^{-22}\text{ eV}$~\cite{2000PhRvL..85.1158H,Marsh:2013ywa,Schive:2014dra,2017PhRvD..95d3541H}.
\item Theoretical uncertainty in the relic density. If the topological defects play a significant role in axion production (i.e. if the PQ symmetry is broken after inflation), then the complex numerical calculations involved in simulating their decay lead to uncertainty in the relic density from different methods.
\item The cosmic axion background. Relativistic axions produced by the decay of a parent will contribute to the ``effective number of neutrinos'', $N_{\rm eff}$, for e.g. cosmic microwave background and BBN constraints. Magnetic fields can also convert these axions into photons, with observable signatures~\cite{2013PhRvL.111o1301C,2013JHEP...10..214C,2014PhRvD..89j3513I}.
\item Thermal axions. If the axion is relativistic when it decouples then it can contribute as hot dark matter. Constraints on hot dark matter are similar to bounds on massive neutrinos, and limit $m_a<0.53\rightarrow 0.62\text{ eV}$ (depending on the analysis) for this population~\cite{2013JCAP...10..020A,2015PhRvD..91l3505D,2016PhLB..752..182D,2005JCAP...07..002H,2007JCAP...08..015H,2008JCAP...04..019H,2010JCAP...08..001H}. For the QCD axion this is not a competitive constraint on $f_a$ compared to bounds from the couplings (see Section~\ref{sec:constraints_coupling}).
\end{itemize}

\subsection{Cosmic Epochs}

Two important epochs define the cosmological evolution of the axion field: PQ symmetry breaking, and the onset of axion field oscillations. The first process is best thought of as thermal (during inflation the distinction is more subtle), while the second process is non-thermal. We recall that during radiation domination the temperature and Hubble rate are related by
\be
H^2M_{pl}^2=\frac{\pi^2}{90}g_{\star,{\rm R}}(T)T^4 \, ,
\label{eqn:friedmann_rad_dom}
\ee
where $g_{\star,{\rm R}}$ is the effective number of relativistic degrees of freedom (a useful analytic fit can be found in the Appendix of Ref.~\cite{Wantz:2009it}). Once $g_\star$ becomes fixed at late times, the temperature simply falls as $1/a$ during the later epochs of matter domination and $\Lambda$ domination. The factor of $M_{pl}$ in Eq.~\eqref{eqn:friedmann_rad_dom} leads to a large hierarchy between $H$ and $T$, separating the scales of thermal and non-thermal phenomena.

\subsubsection{PQ Symmetry Breaking}
\label{sec:two_ics}

Spontaneous symmetry breaking (SSB) occurs when the temperature of the PQ sector falls below the critical temperature, $T_{\rm PQ}<T_c\approx f_a$ (for more details on the thermal field theory, see Ref.~\cite{2017JCAP...08..001B} and references therein). Whether SSB occurs before or after the large scale initial conditions of the Universe were established (for concreteness we will assume inflation, but the same logic applies for other theories) divides axion models into two distinct classes of initial conditions:
\begin{itemize}
\item {\bf Scenario A}: SSB during the ordinary thermal evolution of the Universe.
\item {\bf Scenario B}: SSB before/during inflation (or whatever).
\end{itemize}
The temperature of the PQ sector must be determined. During inflation, the relevant temperature is the Gibbons-Hawking temperature, $T_{\rm GH}=H_I/2\pi$, where $H_I$ is the inflationary Hubble rate.\footnote{More precisely, the Hubble rate when the pivot scale of primordial initial conditions became larger than the horizon.} During the thermal history after inflation, the relevant temperature is that of the standard model. For the QCD axion, the PQ scalar will be in thermal equilibrium with the standard model, mediated by the quarks which couple directly to $\varphi$ prior to the PQ phase transition. For an ALP thermal equilibrium will only be maintained prior to the PQ transition if some standard model particles carry PQ charge. Otherwise the temperatures of the two sectors need not be related.

An important point to note about these scenarios is that \emph{there is a maximum possible temperature of the Universe relevant to Scenario A}, and so all $f_a$ larger than this temperature must be in Scenario B. Table~\ref{tab:i.c._scenario} briefly outlines the differences between these scenarios. 

The full inhomogeneous evolution of the PQ field in Scenario A must be followed in full detail to compute the perturbation spectrum and axion relic density. In principle this is completely determined, though the complexity of the calculation, involving string and domain wall decay, means that computational approximations and assumptions have historically lead to disagreement on this front~\cite{1985PhRvD..32.3172D,1987PhLB..195..361H,1994PhRvL..73.2954B,1996PhRvL..76.2203B}. For some modern calculations, see e.g. Refs.~\cite{2012PhRvD..85j5020H,2017arXiv170807521K}. The small-scale perturbations from SSB have relatively large amplitude and can form gravitationally bound clumps of axions on small scales known as ``miniclusters''~\cite{Hogan:1988mp} with a variety of observational consequences~\cite{1993PhRvL..71.3051K,1994PhRvD..49.5040K,Kolb:1994fi,Kolb:1995bu,2007PhRvD..75d3511Z,2017JHEP...02..046H,Fairbairn:2017dmf,2017PhRvD..95f3017O,2016JCAP...01..035T,Fairbairn:2017sil,Tkachev:2014dpa,Iwazaki:2017rtb,Iwazaki:2014wta}. Some of the most interesting of these minicluster consequences are in gravitational microlensing, signatures in direct detection (including effects on the rate and in the power spectrum), and the possible role of miniclusters as sources of fast radio bursts. Scenario A suffers from a domain wall problem if $N_{\rm DW}>1$~\cite{Sikivie:1982qv}, which disfavours DFSZ type models (the standard DFSZ model has $N_{\rm DW}=6$).

In Scenario B the axion field evolution is much easier to compute thanks to the simplifying power of inflation, which smooths the field, leaving only small amplitude fluctuations that can be evolved using perturbation theory. The smoothing, however, only determines leaves the relative amplitude of axion fluctuations, leaving the overall amplitude as specified by the initial misalignment angle, $\theta_i$, a free parameter. This means that the relic density is also a free parameter, which, depending on $(m_a,f_a,\theta_i)$, can select different regions of parameter space according to your taste for naturalness arguments. For example, the GUT scale QCD axion requires a mild tuning of $\theta_i\approx 10^{-2}$. A constraint on Scenario B emerges from the perturbation spectrum: scale-invariant isocurvature. The amplitude of this spectrum is fixed by $H_I$ and so is directly proportional to the inflationary tensor-to-scalar ratio, $r_T$. CMB constraints on this type of isocuvature (e.g. Refs.~\cite{2009ApJS..180..330K,2016A&A...594A..20P}) imply that most (but not all) axion models in Scenario B are inconsistent with an observably large value of $r_T$~\cite{2008PhRvD..78h3507H,2014PhRvL.113a1801M,Visinelli:2014twa,Hlozek:2017zzf}. An interesting consequence is that in this scenario a measurement of the isocurvature amplitude can be used to measure $H_I$ (if, of course, the existence of axion DM is proven by other means, such as direct detection).

%%%%%%%%%%%%%%%
\begin{table}{r}
\hspace{-0.5in}
\begin{tabular}{|c||c|c|}
\hline
 & Scenario A & Scenario B \\
\hline
\hline
Relic Density  & No free parameters. Complex Calculation. & Simple calculation. Depends on $\theta_i\in [0,\pi]$.  \\
Perturbations$^\star$ &  Small-scale minicluster formation. & Scale-invariant uncorrelated isocurvature.  \\
Notes & Domain wall problem.  & Must occur for large $f_a$.\\
\hline
\end{tabular}
\caption{The main differences between the two scenarios for axion initial conditions. ($^\star$ in addition to the usual scale-invariant adiabatic mode) \label{tab:i.c._scenario}}
\end{table}
%%%%%%%%%%

Let's estimate the maximum value of $f_a$ above which Scenario B must occur in a simple inflationary model. The relevant quantities are the Gibbons-Hawking temperature, and the maximum thermalization temperature after inflation, $T_{\rm max}$ (usually the reheat temperature, though parametric resonance and other dynamics can alter this for the PQ sector). The observational bound on the cosmic microwave background (CMB) tensor-to-scalar ratio, $r_T$, and the measurement of the CMB scalar amplitude $A_s$ imply a bound on $H_I=\pi M_{pl}\sqrt{A_s r_T/2}$, and thus a bound on $T_{\rm GH}<8.2\times 10^{12}\text{ GeV}$ using the results of Refs.~\cite{2016A&A...594A..13P,2016PhRvL.116c1302B}. For simplicity, taking $T_{\rm max}=T_{\rm GH}$, this gives a bound for $f_a>8.2\times 10^{12}\text{ GeV}$ above which Scenario A is excluded. For values of $f_a$ lower than this, whether Scenario A or B occurs is highly model dependent, in particular, either can occur depending on the unknown value of $T_{\rm max}$, which could be as low as Big Bang Nucleosynthesis around 1 MeV.

\subsubsection{Axion Field Evolution}
\label{sec:realignment}
\begin{figure}
\begin{center}
\includegraphics[width=\textwidth]{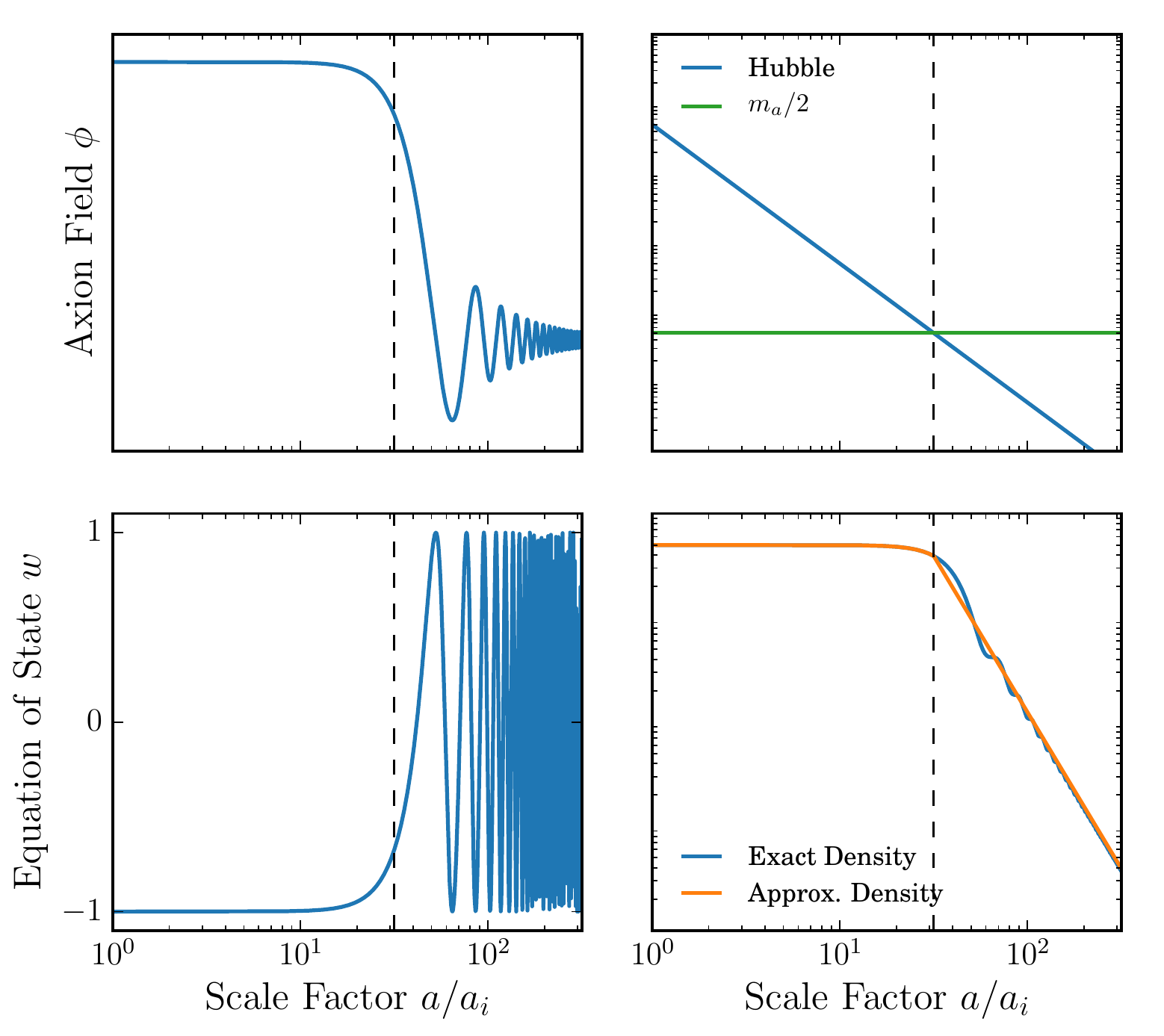}
\caption{Background evolution of the axion field. We show the exact solution for the background evolution in radiation domination for constant axion mass, Eq.~\eqref{eqn:exact_background}. In this simple case, the initial field value and the mass entirely determine the relic density. Reproduced from Ref.~\cite{2016PhR...643....1M}.}
\label{fig:exact_background}
\end{center}
\end{figure}

The second important epoch for axion evolution is the onset of oscillations. The axion equation of motion is:
\be
\Box\phi - \partial_\phi V = 0 \, .
\ee
Taking the homogenous part in a Friedmann-Laamitre-Robertson-Walker background, and expanding the potential to quadratic order:
\be
\ddot{\phi}+3H\dot{\phi}+m_a^2(T)\phi =0\, .
\label{eqn:axion_eom_back}
\ee
For $m_a^2\phi^2\gg 3H\dot{\phi}$ we have $\dot{\phi}\sim m_a\phi$ and so clearly $H\sim m_a$ separates the regimes of overdamped and underdamped motion of the axion field. 

The damping is given by the Friedmann equation:
\be
3 H^2M_{pl}^2=\rho \, ,
\ee
where in general $\rho$ contains all contributions to the energy density, including the axion itself. This is particularly important for dark energy and inflationary axions. For dark matter axions, oscillations must occur during radiation domination when the axion is a sub-dominant component of $\rho$. 

The Hubble rate, $H$, decreases with increasing time (decreasing temperature) while the axion mass increases for increasing time (decreasing temperature). It is customary to define the oscillation temperature implicitly:
\be
3H(T_{\rm osc})=m_a(T_{\rm osc})\, .
\ee

The value of $T_{\rm osc}$ provides a useful reference point (of course we can equally define the scale factor of oscillations, $a_{\rm osc}$, the time of oscillations, etc., which may be more useful in certain cases). For $T\gg T_{\rm osc}$ the axion field is overdamped: the field hardly moves and the energy density contributes to the effective cosmological constant: $\phi\sim {\rm const.}$. For $T\ll T_{\rm osc}$ the axion mass is dominant and the field undergoes damped harmonic motion behaving as non-relativistic matter~\cite{1983PhLB..120..127P,1983PhLB..120..137D,1983PhLB..120..133A,1983PhLB..129...51S,1983PhRvD..28.1243T}: $\phi\sim a^{-3/2}\cos m_a t$. 

From the Friedmann equation in radiation domination, Eq.~\eqref{eqn:friedmann_rad_dom}, we have that $T\propto \sqrt{H M_{pl}}$. Thermally coupled particles become non-relativistic when $T<m$. The largeness of $M_{pl}$, and the non-thermal nature of axion oscillations, however, mean that axions typically begin oscillating for $T\gg m_a$, becoming non-relativistic at a much higher temperature than would a thermally coupled particle of the same mass. This hierarchy between non-thermal and thermal scales explains the relationship between the phenomenology of fuzzy DM and warm DM (see Section~\ref{sec:gravity}).

It is instructive to consider the case of constant axion mass and a radiation dominated background. In that case we have $H\propto 1/2t$ and the exact solution to Eq.~\eqref{eqn:axion_eom_back} is:
\be
\phi=a^{-3/2}(t/t_i)^{1/2}[C_1J_n(m_at)+C_2Y_n(m_at)] \, ,
\label{eqn:exact_background}
\ee
where $n=(3p-1)/2$, $J_n(x)$, $Y_n(x)$ are Bessel functions of the first and second kind, and $t_i$ is the initial time. There is a long-lived attractor solution $\dot{\phi}(t_i)\approx 0$, which we use to fix the ``initial'' conditions via the dimensionful coefficients $C_1$ and $C_2$. This solution displays the limiting behaviours described above, and is shown in Fig.~\ref{fig:exact_background}.

Axion oscillations are a \emph{non-thermal phenomenon}: we are equating the axion mass to the Hubble scale, and we do not actually care what the temperature of the standard model sector is. The evolution of the axion field, and the relic density held its oscillations (see Section~\ref{sec:relic}), are determined entirely by gravitational interactions and coherent axion self interactions independently of their being in thermal contact with the standard model. 

\subsection{The Axion Relic Density}
\label{sec:relic}

The axion energy density is found from the energy-momentum tensor, $T_{\mu\nu}$, with $\rho=-T^0_{\,\,\, 0}$. For the homogeneous component this gives:
\be
\bar{\rho}_a=\frac{1}{2}\dot{\phi}^2+V(\phi) \, .
\ee
We also find the pressure, $3P=T^i_{\,\,\, i}$ giving
\be
\bar{P}_a=\frac{1}{2}\dot{\phi}^2-V(\phi) \, ,
\ee
from which we define the equation of state $w=P/\rho$.

The relic density in axions is defined as the present day energy density relative to the critical density, $\Omega_a h^2=\bar{\rho}(z=0)/3M_H^2M_{pl}^2$, where we have used the reference Hubble rate, $M_H$, defined from $H_0=100 h \text{ km s}^{-1}\text{ Mpc}^{-1}$. In Scenario B the relic density can be computed easily by solving the homogeneous equation of motion for the axion field, a simple ODE. We will discuss Scenario B in detail, and touch on the issues in Scenario A towards the end of this section.

For DM axions in Scenario B the relic density can be estimated using $T_{\rm osc}$. The exact calculation of the relic density in this scenario is:\footnote{Watch out on \url{github.com/doddyphysics} for some example code.}
\begin{itemize}
\item Find $T_{\rm osc}$: $A H(T_{\rm osc})=m_a(T_{\rm osc})$. The choice of $A$ is crucial: more below.
\item Compute the energy density at $T_{\rm osc}$ from the (numerical) solution of the equation of motion for $\theta$ up to this time: $\rho= f_a^2\dot{\theta}^2/2+m_a(T_{\rm osc})^2f_a^2U(\theta)$.
\item Redshift the \emph{number density}, $n_a=\rho_a/m_a$, as non-relativistic matter from this point on (normalising the scale factor to $a(z=0)=1$): $n_a(z=0) = m_a(T_{\rm osc})f_a^2\theta_i^2/2 a(T_{\rm osc})^{3}$. The scale factor can be computed using conservation of entropy (see e.g. Ref.~\cite{1990eaun.book.....K}).
\item Compute the energy density: $\rho(z=0) = n_a(z=0)m_a(T_0)$, where $T_0$ is the temperature today.
\end{itemize}

The first bullet point is key to the accuracy of the calculation. In a full numerical solution we should take $A$ large enough that the field has undergone many oscillations, and that indeed we are in the harmonic regime where $n_a$ is adiabatically conserved. The reason we must make this approximation even for numerical solutions is that for dark matter axions $T_{\rm osc}\gg T_0$ and following a large number of oscillations is numerically prohibitive. As long as $A$ is thus chosen large enough, then with the numerical solution for $\theta$ the other steps are essentially exact.

For \emph{analytic approximations}, we typically take $A=3$ and approximate the axion energy in the second bullet point as being exactly the initial value for a quadratic potential: $\rho_i=m_a(T_{\rm osc})^2f_a^2\theta_i^2/2$. Making these approximations, and using the temperature evolution of the mass consistent with the QCD axion leads to the standard formulae approximating the relic density that can be found in e.g. Refs.~\cite{2004hep.th....9059F,2016PhR...643....1M}.

Obviously the choice of $A$ is related to matching the solutions correctly under this approximation. This is demonstrated in Fig.~\ref{fig:exact_background}, where this approximation works well for $A=2$ for the constant mass ALP in this idealised situation of pure radiation domination for the background. 

In real-Universe examples with a matter-to-radiation transition and late time $\Lambda$ domination, we found in Ref.~\cite{2015PhRvD..91j3512H} that $A=3$ works well for a constant mass ALP with a quadratic potential. In this case, the analytic approximation for the relic density gives \cite{2010PhRvD..82j3528M}:
\begin{eqnarray}
\Omega_a \approx \left\{ 
\begin{array}{ll} 
\frac{1}{6}(9 \Omega_r)^{3/4} \left( \frac{m_a}{H_0} \right)^{1/2} \left\langle\left( \frac{\phi_{i}}{M_{pl}} \right)^2\right\rangle\mbox{if $a_{\rm osc}< a_{\rm eq}$}\, ,\\
\frac{9}{6}\Omega_m \left\langle\left( \frac{\phi_{i}}{M_{pl}} \right)^2\right\rangle\mbox{if $a_{\rm eq}<a_{\rm osc}\lesssim 1$} \, , \\
\frac{1}{6}\left( \frac{m_a}{H_0} \right)^2 \left\langle\left( \frac{\phi_{i}}{M_{pl}} \right)^2\right\rangle\mbox{if $a_{\rm osc}\gtrsim 1$} \, ,\label{eqn:simpledens}
\end{array}
\right. ,
\label{eqn:full_ula_omega}
\end{eqnarray}
where $\phi_i=f_a\theta_i$. The angle brackets appear since in Scenario B the scale-invariant isocurvature perturbations contribute to the mean square misalignment:
\be
\langle \phi^2 \rangle = f_a^2\theta_i^2 + H_I^2/(2\pi)^2 \, .
\ee
In Scenario B the value of $\theta_i$ is a free parameter we can use to set the desired relic density. It is standard to include an ``anharmonic correction factor'', $f_{\rm an}(\theta)$, which provides an additional fudge factor increasing the relic density due to the delayed onset of oscillations when $\theta\sim \pi$. Fits for this can be found in e.g. Ref.~\cite{Diez-Tejedor:2017ivd}, but nothing is a substitute for direct numerical solution.

\begin{figure}
\begin{center}
\includegraphics[width=0.75\textwidth]{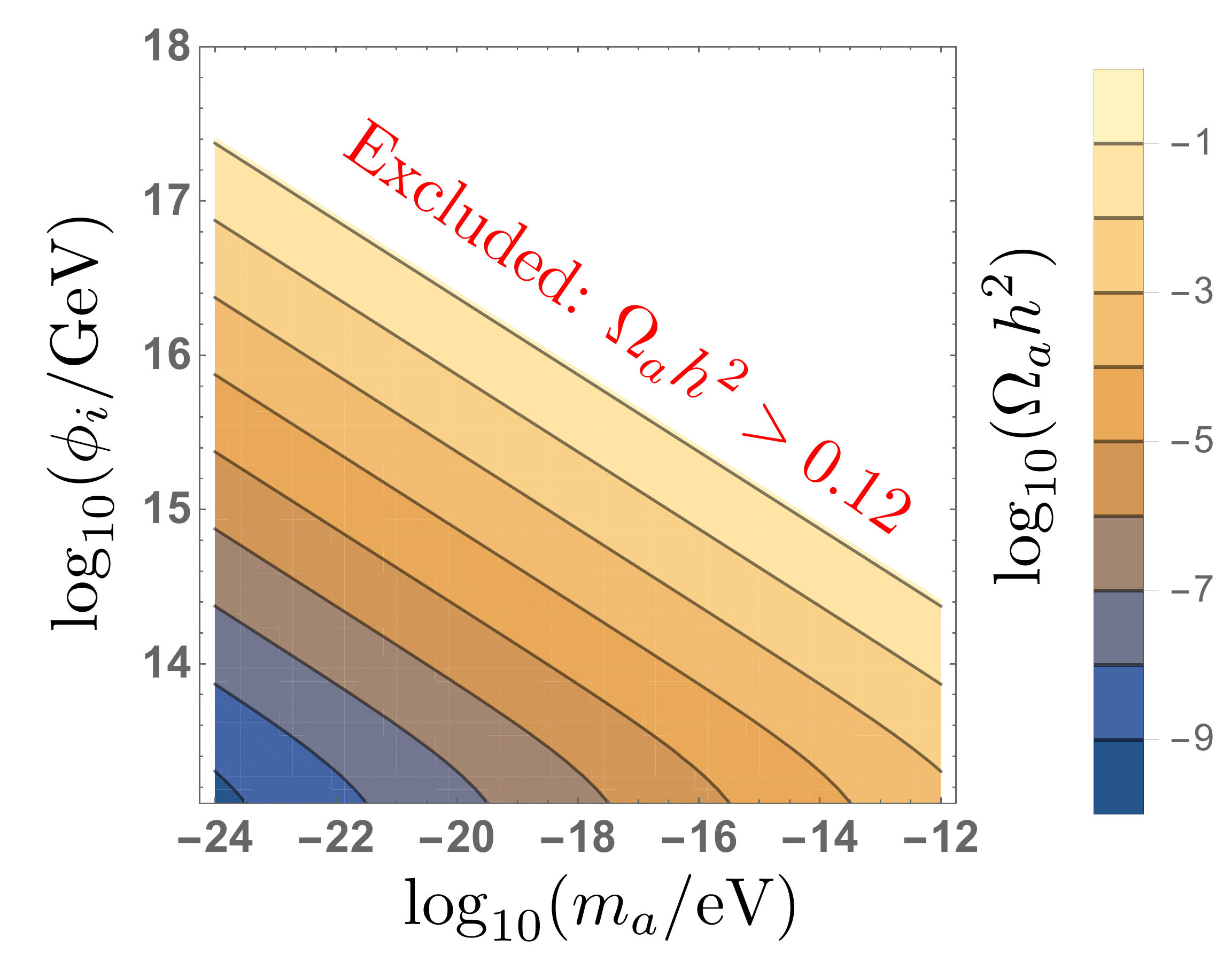}
\caption{ULA relic density from vacuum realignment in Scenario B from the analytic approximation Eq.~\eqref{eqn:full_ula_omega}. Reproduced from Ref.~\cite{2016PhR...643....1M}.}
\label{fig:alp_misalignment}
\end{center}
\end{figure}

Fig.~\ref{fig:alp_misalignment} shows contours of constant relic density from Eq.~\eqref{eqn:full_ula_omega} for ultralight axions (ULAs). We assume a quadratic potential and take with $H_I=10^{14}\text{ GeV}$. 
%%%%%%%%%%%%%
%%%%%%%%%%%%

It is interesting to observe in this plot that values of the decay constant that are natural in a variety of string-inspired models, $f_a\sim 10^{16-17}\text{ GeV}$, provide the correct relic density of axions for masses of the order $m_a=10^{-18-22}\text{ eV}$. This happens to be the mass range of fuzzy DM which is accessible to tests from galaxy formation, and displays interesting signatures that could allow it to be distinguished from standard cold DM (see Section~\ref{sec:gravity}).

\begin{figure}
\begin{center}
\includegraphics[width=0.75\textwidth]{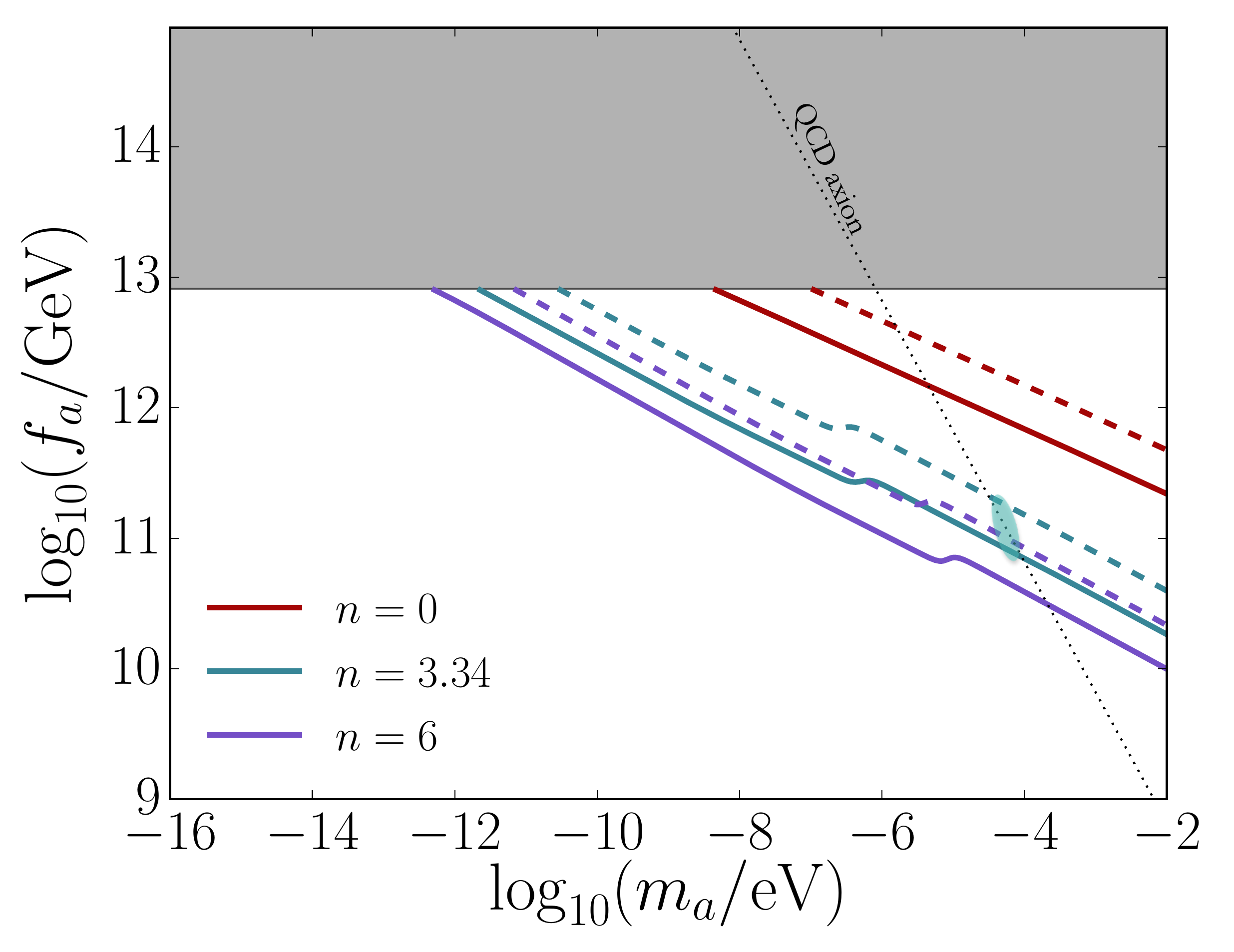}
\caption{Approximation to the relic density in Scenario A for different temperature evolutions of the axion mass parameterised by index $n$. Solid and dashed lines show different values for the effects of anharmonicities and topological defect decay, with the solid lines being the preferred values from Ref.~\cite{2015PhRvD..91f5014K} for defects, and my own fits to anharmonic corrections. Reproduced from Ref.~\cite{Fairbairn:2017sil}.}
\label{fig:relic_contour}
\end{center}
\end{figure}

The relic density computation in Scenario A is far more involved. The full calculation requires solving the inhomogeneous axion equation of motion, which accounts for string and domain wall decay. For $N_{\rm DW}=1$, these effects can be parameterised using a single rescaling of the homogeneous solution by $(1+\alpha_{\rm dec})$, with the simulations of Ref.~\cite{2015PhRvD..91f5014K} favouring $\alpha_{\rm dec.}=2.48$ for the QCD axion. In order to correctly use the rescaling, it is also necessary to use the average value of the homogeneous evolution and any anharmonic corrections:
\be
\langle \theta_i^2f_{\rm an}(\theta_i)\rangle = \frac{1}{2\pi}\int_{-\pi}^{\pi}\theta^2f_{\rm an}(\theta){\rm d}\theta \equiv c_{\rm an}\frac{\pi^2}{3}\, .
\ee
With $\alpha_{\rm dec.}$ and $c_{\rm an.}$ fit from simulations, the relic density can then be computed as in Scenario B, but now with the misalignment angle \emph{fixed} to $\theta_i=\pi/\sqrt{3}$. A more detailed discussion of the calculation can be found in Ref.~\cite{Fairbairn:2017sil}. The results of this approximation are shown in Fig.~\ref{fig:relic_contour}.

%%%%%%%%%%%%%%%%%%%%%%%%%%
\section{Axion Phenomenology}

\subsection{Couplings}
\label{sec:constraints_coupling}
\begin{figure}
\begin{center}
\includegraphics[scale=0.5]{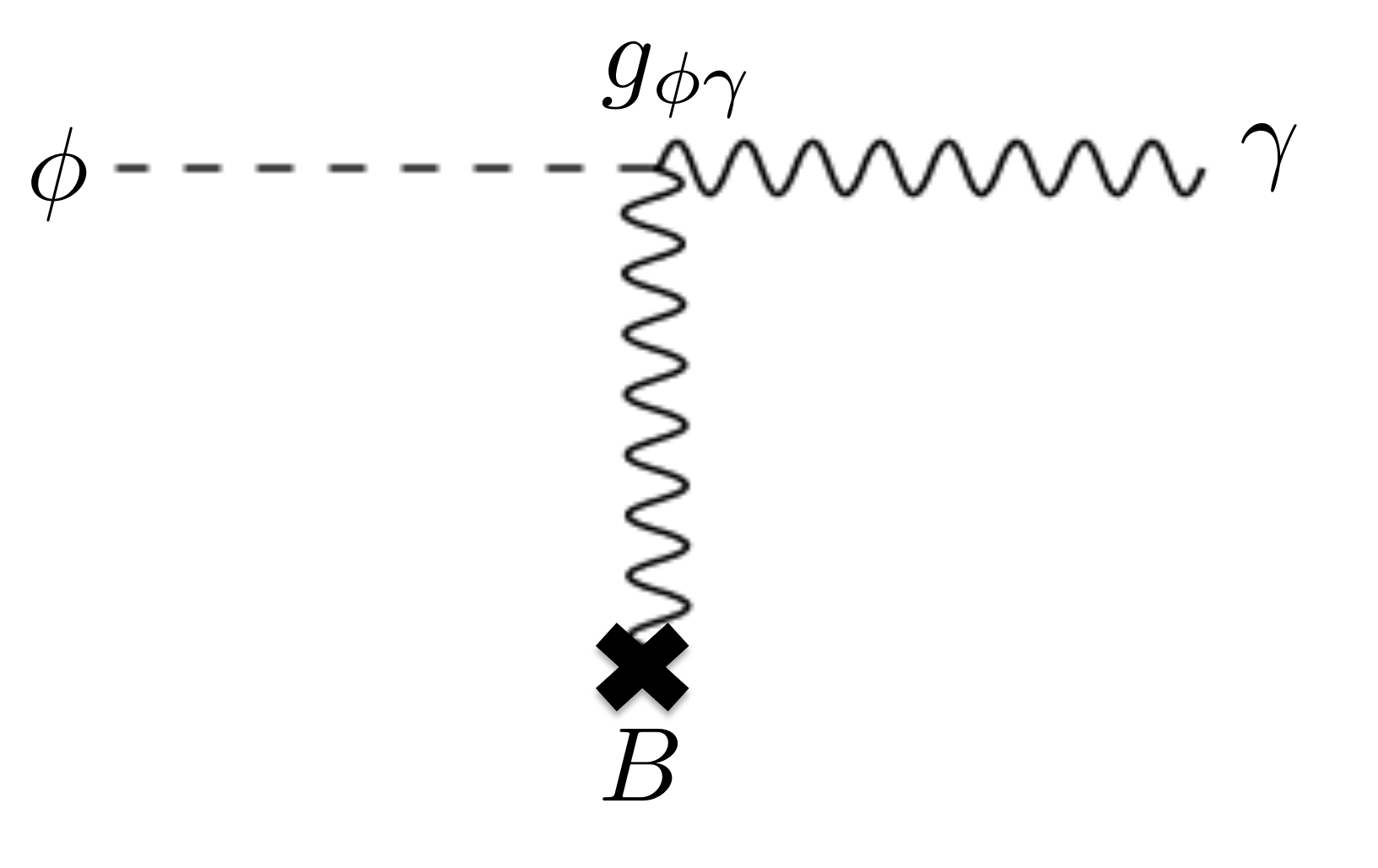}
 \end{center}
 \caption{Axion-photon conversion in the presence of an external magnetic field, $B$. Reproduced from Ref.~\cite{2016PhR...643....1M}.}
\label{fig:primakoff_process}
\end{figure} 

Axion couplings to the standard model have a number of effects that allow the coupling strength to be constrained in the lab and from astrophysics. A thorough review of all experimental constraints on axions is given in Ref.~\cite{2015ARNPS..65..485G}. Global fits are presented in Ref.~\cite{Hoof:2017bqy}. A review of all constraints on the photon coupling is given in Ref.~\cite{2013arXiv1309.7035C}. Briefly, some relevant phenomena are:
\begin{itemize}
\item Stellar evolution. Axions can be produced from standard model particles inside stars. The axions are very weakly interacting and thus easily escape the stars and supernovae, offering an additional cooling channel. The physics and constraints are reviewed by Raffelt in Refs.~\cite{Raffelt:1990yz,2008LNP...741...51R}. A rather robust bound comes from the ratio of horizontal branch stars to red giants found in globular clusters, which bounds the axion photon coupling $g_{\gamma}<6.6\times 10^{-11}\text{ GeV}^{-1}$~\cite{2014PhRvL.113s1302A}. There is also the ``white dwarf cooling hint'' for axions: excess cooling of white dwarfs might be explained by axion emission via the coupling $g_e$~\cite{2012aww..conf..158I}. 
\item Axion mediated forces. The pseudoscalar couplings $g_N$ and $g_e$ mediate a spin-dependent force between standard model particles~\cite{Moody:1984ba}. Constraints on these forces in the laboratory are not very strong compared to the bounds from stellar astrophysics~\cite{2012PhRvD..86a5001R}. However, the proposed ``ARIADNE'' experiment using nuclear magnetic resonance will make substantial improvements, and could even detect the QCD axion for $10^9\text{ GeV}\lesssim f_a\lesssim 10^{12}\text{ GeV}$~\cite{2014PhRvL.113p1801A}.
\item ``Haloscopes'' and other dark matter detection techniques. Using the axion-photon interaction, $g_\gamma$, dark matter axions can be turned into photons in the presence of magnetic fields (see Fig.~\ref{fig:primakoff_process}) inside resonant microwave cavities~\cite{1983PhRvL..51.1415S}. The ADMX experiment is the leader in such constraints~\cite{2010PhRvL.104d1301A}, but many new proposals discussed at this conference will soon also enter the game, including the ADMX high frequency upgrade. Notable new techniques that do not rely on the microwave cavity include the use of resonating circuits~\cite{2014PhRvL.112m1301S} and the ABRACADABRA proposal~\cite{2016PhRvL.117n1801K}; nuclear magnetic resonance and the CASPEr proposal~\cite{2014PhRvX...4b1030B,2017arXiv170705312G}; and dielectric dish antenna and the MADMAX proposal~\cite{TheMADMAXWorkingGroup:2016hpc}. Together, these proposals promise to cover almost the entire parameter space for QCD axion dark matter with $f_a\lesssim 10^{16}\text{ GeV}$. It truly is an exciting time!
\item Axion decays. As noted in Eq.~\eqref{eqn:decay_lifetime} the axion-photon interaction allows axions to decay. Heavy axions decay on cosmological time scales, and are constrained by the effects on the CMB anisotropies, BBN, and CMB spectral distortions~\cite{1992NuPhB.373..399E,1997PhRvD..55.7967M,2015PhRvD..92b3010M}. The strongest constraint comes from the deuterium abundance. Axions and ALPs are generally excluded for masses and lifetimes $1\text{ keV}\lesssim m_a \lesssim 1\text{ GeV}$ and $10^{-4}\text{ s}\lesssim \tau_{\phi\gamma} \lesssim 10^6\text{ s}$
\item Axion-photon conversion in astrophysics. Magnetic fields in clusters convert photons into axions and alter the spectrum of the X-ray photons arriving at Earth. The non-observation of such modulations by the Chandra satellite places a bound on the axion-photon coupling $g_\gamma\lesssim 10^{-12}\text{ GeV}^{-1}$~\cite{2017JCAP...07..005C}. In cosmic magnetic fields the same phenomenon induces CMB spectral distortions, constraining a product of the photon coupling and the cosmic magnetic field strength~\cite{2009JCAP...08..001M,2013PhRvD..88l5024T}. Conversion of axions to photons in the Milky Way magnetic field produces a background of GHz photons correlated with the magnetic field that is accessible to observation by SKA for a range of masses and couplings consistent with QCD axion dark matter~\cite{2017ApJ...845L...4K}, and in the same range that could be detected directly by high frequency ADMX.
\item Anomalous spin precession. The axion coupling to the neutron EDM, $g_d$, and the nucleon coupling, $g_N$ induce spin-precession of neutrons and nuclei in the presence of the axion DM background field. For $g_d$ this occurs in the presence of electric and magnetic fields~\cite{2013PhRvD..88c5023G,2014PhRvX...4b1030B}. For $g_N$ this occurs in magnetic fields, with the axion DM ``wind'' playing the role of a pseudo magnetic field~\cite{2013PhRvD..88c5023G,flambaum_patras}. These effects are the basis of the CASPEr proposal, and have been constrained directly using archival data from nEDM~\cite{Abel:2017rtm}.
\item ``Light Shining Through a Wall''. Axions pass virtually unimpeded through materials (``walls'') that are opaque to photons. Converting a laser photon into an axion using a magnetic field, allowing it to pass through an intermediate wall, and then converting the axion back into a photon, would allow the laser to pass through the wall and indirectly give evidence for axions. This search technique~\cite{2011ConPh..52..211R} is the basis for the ``ALPS'' experiment~\cite{2010PhLB..689..149E,2013JInst...8.9001B}, which aims at constraining $g_{\phi\gamma}\sim 2\times 10^{-11}\text{ GeV}^{-1}$ in the upgraded version currently in operation.
\item ``Helioscopes''. Axions from the sun can be converted into visible photons inside a telescope with a magnetic field~\cite{1983PhRvL..51.1415S}. Constraints on $g_\gamma\lesssim 10^{-10}\text{ GeV}^{-1}$ using this technique have been presented by the CAST experiment~\cite{2005PhRvL..94l1301Z,2007JCAP...04..010A,2014PhRvL.112i1302A}. The proposed ``International Axion Observatory'' would improve the limits by an order of magnitude~\cite{Vogel:2013bta}.
\end{itemize} 

\subsection{Gravitation}
\label{sec:gravity}

The classical axion field, $\phi$, has novel gravitational effects caused by the Compton wavelength, and by the axion potential.
\begin{itemize}
\item ``Fuzzy'' Dark Matter. As already discussed, when the axion mass is very small, $m_a\sim 10^{-22}\text{ eV}$, the field displays coherence on astrophysical length scales~\cite{2000PhRvL..85.1158H,Marsh:2013ywa,Schive:2014dra,2017PhRvD..95d3541H}. This makes ULAs distinct from cold DM, and drives the lower bound on DM particle mass from cosmological constraints such as the CMB anisotropies~\cite{2006PhLB..642..192A,2015PhRvD..91j3512H,Hlozek:2017zzf}, the high redshift luminosity function~\cite{2015MNRAS.450..209B,2016ApJ...818...89S,2017PhRvD..95h3512C}, and the Lyman-$\alpha$ forest flux power spectrum~\cite{2017PhRvL.119c1302I,2017MNRAS.471.4606A}. Superficially the model resembles warm DM~\cite{bode2001}, suppressing cosmic structure formation below a certain length scale, and with the warm DM mass, $m_X$, roughly related to the axion mass as $m_X\sim \sqrt{m_aM_{pl}}$. However, the small scale physics is very different and fuzzy DM requires dedicated simulations, which reveal striking unique features such as soliton formation, and quasi-particles. A number of beyond-CDM simulations and semi-analytic methods have been developed to study this novel type of DM~\cite{Schive:2014dra,2016PhRvD..94l3523V,2017MNRAS.465..941D,2016PhRvD..94d3513S,2017MNRAS.471.4559M}. 
\item Black hole superradiance. This gravitational phenomenon applies to all bosonic fields, with different timescales depending on the spin (zero, one, or two). A population of bosons is built up in a ``gravitational atom'' around the black hole \emph{from vacuum fluctuations}. Thus this phenomenon makes no assumptions about the cosmic density or origin of the bosonic field. Spin is extracted from the black hole via the Penrose process~\cite{1969NCimR...1..252P}. The boson mass provides a potential barrier (a ``mirror''), and the process becomes runaway~\cite{1972Natur.238..211P,1973ApJ...185..649P}. The resulting spin down of black holes makes certain regions on the ``Regge plane'' (mass versus spin plane) effectively forbidden for astrophysical black holes. Astrophysical observations of rapidly rotating black holes thus exclude bosons of certain ranges of mass~\cite{2015PhRvD..91h4011A,axiverse,2011PhRvD..83d4026A,2012PhRvL.109m1102P,2015CQGra..32m4001B}. For the spin zero axion, stellar mass black holes exclude $6\times 10^{-13}\text{ eV}<m_a<2\times 10^{-11}\text{ eV}$ at $2\sigma$, which for the QCD axion excludes $3\times 10^{17}\text{ GeV}<f_a<1\times 10^{19}\text{ GeV}$. The supermassive BH measurements give only $1\sigma$ exclusions $10^{-18}\text{ eV}<m_a<10^{-16}\text{ eV}$. The energy extracted from the BH angular momentum can be emitted by the axion cloud in the form of gravitational waves (GWs). The recent direct detection of GWs by LIGO~\cite{2016PhRvL.116f1102A} opens up an exciting new opportunity to study axions and other light bosons from the inferred mass and spin distributions of BHs, and from direct GW signals of superradiance~\cite{2017PhRvD..95d3001A,2017PhRvD..96f4050B}.
\item Oscillating dark matter pressure. The axion equation of state (shown in Fig.~\ref{fig:exact_background}) oscillates with frequency $2m_a$, originating from an oscillating pressure. The axion is only pressureless when averaged over time scales larger than $(2m_a)^{-1}$. Pressure oscillations induce oscillations in the metric potentials, which manifest as a scalar strain on pulsar timing arrays (PTAs) and in gravitational wave detectors (just as gravitational waves are a tensor strain)~\cite{2014JCAP...02..019K}. The NANOGrav PTA sets limits an order of magnitude higher than the expected signal at $m_a=10^{-23}\text{ eV}$~\cite{2014PhRvD..90f2008P}, though SKA is forecast to be sensitive to the signal at this mass even if such ULAs constitute just 1\% of the DM~\cite{2014JCAP...02..019K}. In GW detectors, the axion DM wind anisotropic stress, $\sigma_{ij}\propto \nabla_i\phi\nabla_j\phi$, manifests as a ``scalar GW'' also~\cite{2017IJMPD..2650063A}. 
\item Axion stars. The gradient term in the Klein Gordon equation opposes gravitational collapse of scalar fields on small scales. This leads to the existence of a class of pseudo-solitonic boson star known as an oscilloton~\cite{1969PhRv..187.1767R,1991PhRvL..66.1659S} for the case of a massive real scalar. For axions, these solutions are ``axion stars''. These objects are formed during gravitational collapse, halted by the effective pressure of the gradients. Emission of scalar waves leads to ``gravitational cooling''~\cite{1994PhRvL..72.2516S}, and the stars settle into the ground state. The solitons are a condensate of coherent axions. Axions stars are observed to form in the centres of  DM halos in numerical simulations~\cite{Schive:2014dra,Schive:2014hza,2016PhRvD..94d3513S}. This density core may play a role in the presence of cores in dwarf spheroidal satellites of the Milky Way~\cite{2015MNRAS.451.2479M,2017MNRAS.472.1346G,2017MNRAS.468.1338C}. Axion stars should also be present in the centre of miniclusters, and, if axion self-scattering is efficient, entire miniclusters might thus condense~\cite{1993PhRvL..71.3051K}. Axion stars have a maximum mass above which they become unstable~\cite{2017JCAP...03..055H}. For weak self-interactions, $f_a\gtrsim M_{pl}$, the instability leads to black hole formation, while for stronger self-interactions the instability results in emission of relativistic axion waves~\cite{2017JCAP...03..055H,2017PhRvL.118a1301L,2016PhRvD..94h3007C,2016MPLA...3150090E,2017arXiv171006268C,2016JHEP...12..066E}. Axion stars could detected as they pass through the Earth using a network of magnetometers~\cite{2017arXiv171004323K}. Especially compact axion stars could lead to unique signatures in gravitational wave detectors from their binary inspirals with each other and with other astrophysical objects~\cite{2016JCAP...10..001G}. 
\item Inflation and Dark Energy. The periodic nature of the axion potential implies that there are maxima where the potential is locally flat. The (tachyonic) mass at the maximum is protected from perturbative quantum corrections by the shift symmetry, leading to fairly natural models for inflation~\cite{1990PhRvL..65.3233F} and dark energy~\cite{1995PhRvL..75.2077F}. The axion potential contributes to the effective cosmological constant. If the field is placed sufficiently close to the top of the potential, then a sufficient number of $e$-foldings of inflation can be driven (for dark energy the requirement is instead on the equation of state). Constraints on axion dark energy can be found in Ref.~\cite{2017JCAP...01..023S}. The simplest model of natural inflation takes $V(\phi)\propto \cos (\phi/f_a)$. After normalising the scalar CMB amplitude, this model is a two parameter family giving predicting a strip in the plane of scalar spectral index versus tensor-to-scalar ratio, $(n_s,r_T)$. It is consistent with the \emph{Planck} results~\cite{planck2015}, but could could be excluded by CMB-S4~\cite{2016arXiv161002743A}. Variants on axion inflation inspired by string theory are $N$-flation~\cite{2008JCAP...08..003D} and axion monodromy~\cite{2008PhRvD..78j6003S,2010PhRvD..82d6003M}. Both models seek to deal with issues relating to super-Planckian field excursions, the Lyth bound for $r_T$~\cite{1997PhRvL..78.1861L}, and the ``weak gravity conjecture''~\cite{2007JHEP...06..060A}, and construct string-inspired models with observably large $r_T$. 
\end{itemize}

\section*{Acknowledgements}

These brief notes contributed to the 13th Patras Workshop on Axions, WIMPs and WISPs, Thessaloniki, May 15 to 19, 2017. They rely heavily on the review Ref.~\cite{2016PhR...643....1M}, which is far more complete, and from which most figures are taken (though of course these notes discuss new results of the last two years). I am grateful to Thomas Bachlechner, Laura Covi, Jihn Kim, Chris Murphy, Maxim Pospelov, and Andreas Ringwald for useful discussions, and to Vishnu Jejala for providing the data for Fig.~\ref{fig:kreuzer_skarke}. Work at King's College London was supported by a Royal Astronomical Society Postdoctoral Fellowship. Research at the University of G\"{o}ttingen is funded by the Alexander von Humboldt Foundation and the German Federal Ministry of Education and Research.

\begin{footnotesize}

\bibliographystyle{h-physrev3.bst}
\bibliography{axion_review}

\begin{thebibliography}{100}

\bibitem{pecceiquinn1977}
R.~Peccei and H.~R. Quinn,
\newblock \prl {\bf 38}, 1440 (1977).
%%CITATION = PRLTA,38,1440;%%

\bibitem{weinberg1978}
S.~Weinberg,
\newblock \prl {\bf 40}, 223 (1978).
%%CITATION = PRLTA,40,223;%%

\bibitem{wilczek1978}
F.~Wilczek,
\newblock \prl {\bf 40}, 279 (1978).
%%CITATION = PRLTA,40,279;%%

\bibitem{1979PhLB...88..123C}
R.~J. {Crewther}, P.~{di Vecchia}, G.~{Veneziano}, and E.~{Witten},
\newblock Phys. Lett. B {\bf 88}, 123 (1979).

\bibitem{2015PhRvD..92i2003P}
J.~M. {Pendlebury} {\em et~al.},
\newblock \prd {\bf 92}, 092003 (2015), 1509.04411.

\bibitem{2014ChPhC..38i0001O}
K.~A. {Olive} and {Particle Data Group},
\newblock Chinese Physics C {\bf 38}, 090001 (2014), 1412.1408.

\bibitem{Srednicki:1985xd}
M.~Srednicki,
\newblock Nucl. Phys. {\bf B260}, 689 (1985).
%%CITATION = NUPHA,B260,689;%%

\bibitem{2003qftn.book.....Z}
A.~{Zee},
\newblock {\em {Quantum field theory in a nutshell}} (, 2003).

\bibitem{2016PhR...643....1M}
D.~J.~E. {Marsh},
\newblock \physrep {\bf 643}, 1 (2016), 1510.07633.

\bibitem{1988assy.book.....C}
S.~{Coleman},
\newblock {\em {Aspects of Symmetry}} (Cambridge University Press, 1988).

\bibitem{Vafa:1984xg}
C.~Vafa and E.~Witten,
\newblock \prl {\bf 53}, 535 (1984).
%%CITATION = PRLTA,53,535;%%

\bibitem{2016JHEP...01..034D}
G.~G. {di Cortona}, E.~{Hardy}, J.~P. {Vega}, and G.~{Villadoro},
\newblock Journal of High Energy Physics {\bf 1}, 34 (2016), 1511.02867.

\bibitem{1981RvMP...53...43G}
D.~J. {Gross}, R.~D. {Pisarski}, and L.~G. {Yaffe},
\newblock Reviews of Modern Physics {\bf 53}, 43 (1981).

\bibitem{2016PhLB..752..175B}
S.~{Borsanyi} {\em et~al.},
\newblock Physics Letters B {\bf 752}, 175 (2016), 1508.06917.

\bibitem{Borsanyi:2016ksw}
S.~Borsanyi {\em et~al.},
\newblock Nature {\bf 539}, 69 (2016), 1606.07494.
%%CITATION = ARXIV:1606.07494;%%

\bibitem{Sikivie:1982qv}
P.~Sikivie,
\newblock \prl {\bf 48}, 1156 (1982).
%%CITATION = PRLTA,48,1156;%%

\bibitem{Kim:1986ax}
J.~E. Kim,
\newblock Phys. Rept. {\bf 150}, 1 (1987).
%%CITATION = PRPLC,150,1;%%

\bibitem{2015PhRvD..92b3010M}
M.~{Millea}, L.~{Knox}, and B.~D. {Fields},
\newblock \prd {\bf 92}, 023010 (2015), 1501.04097.

\bibitem{1979PhRvL..43..103K}
J.~E. {Kim},
\newblock \prl {\bf 43}, 103 (1979).

\bibitem{1980NuPhB.166..493S}
M.~A. {Shifman}, A.~I. {Vainshtein}, and V.~I. {Zakharov},
\newblock Nuclear Physics B {\bf 166}, 493 (1980).

\bibitem{1981PhLB..104..199D}
M.~{Dine}, W.~{Fischler}, and M.~{Srednicki},
\newblock Phys. Lett. B {\bf 104}, 199 (1981).

\bibitem{Zhitnitsky:1980tq}
A.~Zhitnitsky,
\newblock Sov.J . Nucl. Phys. {\bf 31}, 260 (1980).
%%CITATION = SJNCA,31,260;%%

\bibitem{2017PhRvL.118c1801D}
L.~{Di Luzio}, F.~{Mescia}, and E.~{Nardi},
\newblock Physical Review Letters {\bf 118}, 031801 (2017), 1610.07593.

\bibitem{2006JHEP...06..051S}
P.~{Svrcek} and E.~{Witten},
\newblock \jhep {\bf 6}, 51 (2006), hep-th/0605206.

\bibitem{2006JHEP...05..078C}
J.~P. {Conlon},
\newblock Journal of High Energy Physics {\bf 5}, 078 (2006), hep-th/0602233.

\bibitem{2007stmt.book.....B}
K.~{Becker}, M.~{Becker}, and J.~H. {Schwarz},
\newblock {\em {String Theory and M-Theory}} (Cambridge University Press,
  2007).

\bibitem{2012arXiv1209.2299R}
A.~{Ringwald},
\newblock ArXiv e-prints  (2012), 1209.2299.

\bibitem{2012JHEP...10..146C}
M.~{Cicoli}, M.~D. {Goodsell}, and A.~{Ringwald},
\newblock \jhep {\bf 10}, 146 (2012), 1206.0819.

\bibitem{2004sgig.book.....C}
S.~M. {Carroll},
\newblock {\em {Spacetime and geometry. An introduction to general relativity}}
  (Addison Wesley, 2004).

\bibitem{1987cup..bookR....G}
M.~B. {Green}, J.~H. {Schwarz}, and E.~{Witten},
\newblock {\em {Superstring theory. Volume 2 - Loop amplitudes, anomalies and
  phenomenology}} (Cambridge University Press, 1987).

\bibitem{1985NuPhB.258...46C}
P.~{Candelas}, G.~T. {Horowitz}, A.~{Strominger}, and E.~{Witten},
\newblock Nuclear Physics B {\bf 258}, 46 (1985).

\bibitem{Altman:2014bfa}
R.~Altman, J.~Gray, Y.-H. He, V.~Jejjala, and B.~D. Nelson,
\newblock JHEP {\bf 02}, 158 (2015), 1411.1418.
%%CITATION = ARXIV:1411.1418;%%

\bibitem{He:2017aed}
Y.-H. He,
\newblock (2017), 1706.02714.
%%CITATION = ARXIV:1706.02714;%%

\bibitem{Kreuzer:2000xy}
M.~Kreuzer and H.~Skarke,
\newblock Adv. Theor. Math. Phys. {\bf 4}, 1209 (2002), hep-th/0002240.
%%CITATION = HEP-TH/0002240;%%

\bibitem{2000math......1106K}
M.~{Kreuzer} and H.~{Skarke},
\newblock ArXiv Mathematics e-prints  (2000), math/0001106.

\bibitem{Grimm:2004uq}
T.~W. Grimm and J.~Louis,
\newblock Nucl. Phys. {\bf B699}, 387 (2004), hep-th/0403067.
%%CITATION = HEP-TH/0403067;%%

\bibitem{axiverse}
A.~{Arvanitaki}, S.~{Dimopoulos}, S.~{Dubovsky}, N.~{Kaloper}, and
  J.~{March-Russell},
\newblock \prd {\bf 81}, 123530 (2010), 0905.4720.

\bibitem{2006JCAP...05..018E}
R.~{Easther} and L.~{McAllister},
\newblock \jcap {\bf 5}, 018 (2006), hep-th/0512102.

\bibitem{2017PhRvD..96h3510S}
M.~J. {Stott}, D.~J.~E. {Marsh}, C.~{Pongkitivanichkul}, L.~C. {Price}, and
  B.~S. {Acharya},
\newblock \prd {\bf 96}, 083510 (2017), 1706.03236.

\bibitem{Bachlechner:2017zpb}
T.~C. Bachlechner, K.~Eckerle, O.~Janssen, and M.~Kleban,
\newblock (2017), 1703.00453.
%%CITATION = ARXIV:1703.00453;%%

\bibitem{1988NuPhB.306..890G}
S.~B. {Giddings} and A.~{Strominger},
\newblock Nuclear Physics B {\bf 306}, 890 (1988).

\bibitem{2017arXiv170607415A}
R.~{Alonso} and A.~{Urbano},
\newblock ArXiv e-prints  (2017), 1706.07415.

\bibitem{2013PhRvD..88c5023G}
P.~W. {Graham} and S.~{Rajendran},
\newblock \prd {\bf 88}, 035023 (2013), 1306.6088.

\bibitem{2014JHEP...06..037D}
A.~G. {Dias}, A.~C.~B. {Machado}, C.~C. {Nishi}, A.~{Ringwald}, and
  P.~{Vaudrevange},
\newblock Journal of High Energy Physics {\bf 6}, 37 (2014), 1403.5760.

\bibitem{2016PhRvD..93b5027K}
J.~E. {Kim} and D.~J.~E. {Marsh},
\newblock \prd {\bf 93}, 025027 (2016), 1510.01701.

\bibitem{2008PhRvD..78j6003S}
E.~{Silverstein} and A.~{Westphal},
\newblock \prd {\bf 78}, 106003 (2008), 0803.3085.

\bibitem{2010PhRvD..82d6003M}
L.~{McAllister}, E.~{Silverstein}, and A.~{Westphal},
\newblock \prd {\bf 82}, 046003 (2010), 0808.0706.

\bibitem{1992PhLB..282..137K}
M.~{Kamionkowski} and J.~{March-Russell},
\newblock Physics Letters B {\bf 282}, 137 (1992), hep-th/9202003.

\bibitem{1995PhRvD..52..912K}
R.~{Kallosh}, A.~{Linde}, D.~{Linde}, and L.~{Susskind},
\newblock \prd {\bf 52}, 912 (1995), hep-th/9502069.

\bibitem{2000PhRvL..85.1158H}
W.~{Hu}, R.~{Barkana}, and A.~{Gruzinov},
\newblock Physical Review Letters {\bf 85}, 1158 (2000), astro-ph/0003365.

\bibitem{Marsh:2013ywa}
D.~J.~E. {Marsh} and J.~{Silk},
\newblock \mnras {\bf 437}, 2652 (2014), 1307.1705.

\bibitem{Schive:2014dra}
H.-Y. {Schive}, T.~{Chiueh}, and T.~{Broadhurst},
\newblock Nature Physics {\bf 10}, 496 (2014), 1406.6586.

\bibitem{2017PhRvD..95d3541H}
L.~{Hui}, J.~P. {Ostriker}, S.~{Tremaine}, and E.~{Witten},
\newblock \prd {\bf 95}, 043541 (2017), 1610.08297.

\bibitem{2013PhRvL.111o1301C}
J.~P. {Conlon} and M.~C.~D. {Marsh},
\newblock \prl {\bf 111}, 151301 (2013), 1305.3603.

\bibitem{2013JHEP...10..214C}
J.~P. {Conlon} and M.~C.~D. {Marsh},
\newblock \jhep {\bf 10}, 214 (2013), 1304.1804.

\bibitem{2014PhRvD..89j3513I}
L.~{Iliesiu}, D.~J.~E. {Marsh}, K.~{Moodley}, and S.~{Watson},
\newblock \prd {\bf 89}, 103513 (2014), 1312.3636.

\bibitem{2013JCAP...10..020A}
M.~{Archidiacono}, S.~{Hannestad}, A.~{Mirizzi}, G.~{Raffelt}, and Y.~Y.~Y.
  {Wong},
\newblock \jcap {\bf 10}, 020 (2013), 1307.0615.

\bibitem{2015PhRvD..91l3505D}
E.~{Di Valentino}, S.~{Gariazzo}, E.~{Giusarma}, and O.~{Mena},
\newblock \prd {\bf 91}, 123505 (2015), 1503.00911.

\bibitem{2016PhLB..752..182D}
E.~{Di Valentino} {\em et~al.},
\newblock Physics Letters B {\bf 752}, 182 (2016), 1507.08665.

\bibitem{2005JCAP...07..002H}
S.~{Hannestad}, A.~{Mirizzi}, and G.~{Raffelt},
\newblock \jcap {\bf 7}, 002 (2005), hep-ph/0504059.

\bibitem{2007JCAP...08..015H}
S.~{Hannestad}, A.~{Mirizzi}, G.~G. {Raffelt}, and Y.~Y.~Y. {Wong},
\newblock \jcap {\bf 8}, 015 (2007), 0706.4198.

\bibitem{2008JCAP...04..019H}
S.~{Hannestad}, A.~{Mirizzi}, G.~G. {Raffelt}, and Y.~Y.~Y. {Wong},
\newblock \jcap {\bf 4}, 019 (2008), 0803.1585.

\bibitem{2010JCAP...08..001H}
S.~{Hannestad}, A.~{Mirizzi}, G.~G. {Raffelt}, and Y.~Y.~Y. {Wong},
\newblock \jcap {\bf 8}, 001 (2010), 1004.0695.

\bibitem{Wantz:2009it}
O.~Wantz and E.~Shellard,
\newblock \prd {\bf 82}, 123508 (2010), 0910.1066.
%%CITATION = ARXIV:0910.1066;%%

\bibitem{2017JCAP...08..001B}
G.~{Ballesteros}, J.~{Redondo}, A.~{Ringwald}, and C.~{Tamarit},
\newblock \jcap {\bf 8}, 001 (2017), 1610.01639.

\bibitem{1985PhRvD..32.3172D}
R.~L. {Davis},
\newblock \prd {\bf 32}, 3172 (1985).

\bibitem{1987PhLB..195..361H}
D.~{Harari} and P.~{Sikivie},
\newblock Phys. Lett. B {\bf 195}, 361 (1987).

\bibitem{1994PhRvL..73.2954B}
R.~A. {Battye} and E.~P.~S. {Shellard},
\newblock \prl {\bf 73}, 2954 (1994), astro-ph/9403018.

\bibitem{1996PhRvL..76.2203B}
R.~A. {Battye} and E.~P.~S. {Shellard},
\newblock \prl {\bf 76}, 2203 (1996).

\bibitem{2012PhRvD..85j5020H}
T.~{Hiramatsu}, M.~{Kawasaki}, K.~{Saikawa}, and T.~{Sekiguchi},
\newblock \prd {\bf 85}, 105020 (2012), 1202.5851.

\bibitem{2017arXiv170807521K}
V.~B. {Klaer} and G.~D. {Moore},
\newblock ArXiv e-prints  (2017), 1708.07521.

\bibitem{Hogan:1988mp}
C.~J. {Hogan} and M.~J. {Rees},
\newblock Phys. Lett. B {\bf 205}, 228 (1988).

\bibitem{1993PhRvL..71.3051K}
E.~W. {Kolb} and I.~I. {Tkachev},
\newblock \prl {\bf 71}, 3051 (1993), hep-ph/9303313.

\bibitem{1994PhRvD..49.5040K}
E.~W. {Kolb} and I.~I. {Tkachev},
\newblock \prd {\bf 49}, 5040 (1994), astro-ph/9311037.

\bibitem{Kolb:1994fi}
E.~W. {Kolb} and I.~I. {Tkachev},
\newblock \prd {\bf 50}, 769 (1994), astro-ph/9403011.

\bibitem{Kolb:1995bu}
E.~W. {Kolb} and I.~I. {Tkachev},
\newblock \apjl {\bf 460}, L25 (1996), astro-ph/9510043.

\bibitem{2007PhRvD..75d3511Z}
K.~M. {Zurek}, C.~J. {Hogan}, and T.~R. {Quinn},
\newblock \prd {\bf 75}, 043511 (2007), astro-ph/0607341.

\bibitem{2017JHEP...02..046H}
E.~{Hardy},
\newblock Journal of High Energy Physics {\bf 2}, 46 (2017), 1609.00208.

\bibitem{Fairbairn:2017dmf}
M.~Fairbairn, D.~J.~E. Marsh, and J.~Quevillon,
\newblock Phys. Rev. Lett. {\bf 119}, 021101 (2017), 1701.04787.
%%CITATION = ARXIV:1701.04787;%%

\bibitem{2017PhRvD..95f3017O}
C.~A.~J. {O'Hare} and A.~M. {Green},
\newblock \prd {\bf 95}, 063017 (2017), 1701.03118.

\bibitem{2016JCAP...01..035T}
P.~{Tinyakov}, I.~{Tkachev}, and K.~{Zioutas},
\newblock \jcap {\bf 1}, 035 (2016), 1512.02884.

\bibitem{Fairbairn:2017sil}
M.~Fairbairn, D.~J.~E. Marsh, J.~Quevillon, and S.~Rozier,
\newblock (2017), 1707.03310.
%%CITATION = ARXIV:1707.03310;%%

\bibitem{Tkachev:2014dpa}
I.~I. Tkachev,
\newblock JETP Lett. {\bf 101}, 1 (2015), 1411.3900,
\newblock [Pisma Zh. Eksp. Teor. Fiz.101,no.1,3(2015)].
%%CITATION = ARXIV:1411.3900;%%

\bibitem{Iwazaki:2017rtb}
A.~Iwazaki,
\newblock (2017), 1707.04827.
%%CITATION = ARXIV:1707.04827;%%

\bibitem{Iwazaki:2014wta}
A.~Iwazaki,
\newblock (2014), 1412.7825.
%%CITATION = ARXIV:1412.7825;%%

\bibitem{2009ApJS..180..330K}
E.~{Komatsu} {\em et~al.},
\newblock \apjs {\bf 180}, 330 (2009), 0803.0547.

\bibitem{2016A&A...594A..20P}
{Planck Collaboration} {\em et~al.},
\newblock \aap {\bf 594}, A20 (2016), 1502.02114.

\bibitem{2008PhRvD..78h3507H}
M.~P. {Hertzberg}, M.~{Tegmark}, and F.~{Wilczek},
\newblock \prd {\bf 78}, 083507 (2008), 0807.1726.

\bibitem{2014PhRvL.113a1801M}
D.~J.~E. {Marsh}, D.~{Grin}, R.~{Hlozek}, and P.~G. {Ferreira},
\newblock Physical Review Letters {\bf 113}, 011801 (2014).

\bibitem{Visinelli:2014twa}
L.~Visinelli and P.~Gondolo,
\newblock \prl {\bf 113}, 011802 (2014), 1403.4594.
%%CITATION = ARXIV:1403.4594;%%

\bibitem{Hlozek:2017zzf}
R.~Hlozek, D.~J.~E. Marsh, and D.~Grin,
\newblock (2017), 1708.05681.
%%CITATION = ARXIV:1708.05681;%%

\bibitem{2016A&A...594A..13P}
{Planck Collaboration} {\em et~al.},
\newblock \aap {\bf 594}, A13 (2016), 1502.01589.

\bibitem{2016PhRvL.116c1302B}
{BICEP2 Collaboration} {\em et~al.},
\newblock \prl {\bf 116}, 031302 (2016), 1510.09217.

\bibitem{1983PhLB..120..127P}
J.~{Preskill}, M.~B. {Wise}, and F.~{Wilczek},
\newblock Phys. Lett. B {\bf 120}, 127 (1983).

\bibitem{1983PhLB..120..137D}
M.~{Dine} and W.~{Fischler},
\newblock Phys. Lett. B {\bf 120}, 137 (1983).

\bibitem{1983PhLB..120..133A}
L.~F. {Abbott} and P.~{Sikivie},
\newblock Phys. Lett. B {\bf 120}, 133 (1983).

\bibitem{1983PhLB..129...51S}
P.~J. {Steinhardt} and M.~S. {Turner},
\newblock Phys. Lett. B {\bf 129}, 51 (1983).

\bibitem{1983PhRvD..28.1243T}
M.~S. {Turner},
\newblock \prd {\bf 28}, 1243 (1983).

\bibitem{1990eaun.book.....K}
E.~W. {Kolb} and M.~S. {Turner},
\newblock {\em {The early universe.}} (Addison-Wesley, 1990).

\bibitem{2004hep.th....9059F}
P.~{Fox}, A.~{Pierce}, and S.~{Thomas},
\newblock ArXiv High Energy Physics - Theory e-prints  (2004), hep-th/0409059.

\bibitem{2015PhRvD..91j3512H}
R.~{Hlo\v{z}ek}, D.~{Grin}, D.~J.~E. {Marsh}, and P.~G. {Ferreira},
\newblock \prd {\bf 91}, 103512 (2015), 1410.2896.

\bibitem{2010PhRvD..82j3528M}
D.~J.~E. {Marsh} and P.~G. {Ferreira},
\newblock \prd {\bf 82}, 103528 (2010), 1009.3501.

\bibitem{Diez-Tejedor:2017ivd}
A.~Diez-Tejedor and D.~J.~E. Marsh,
\newblock (2017), 1702.02116.
%%CITATION = ARXIV:1702.02116;%%

\bibitem{2015PhRvD..91f5014K}
M.~{Kawasaki}, K.~{Saikawa}, and T.~{Sekiguchi},
\newblock \prd {\bf 91}, 065014 (2015), 1412.0789.

\bibitem{2015ARNPS..65..485G}
P.~W. {Graham}, I.~G. {Irastorza}, S.~K. {Lamoreaux}, A.~{Lindner}, and K.~A.
  {van Bibber},
\newblock Annual Review of Nuclear and Particle Science {\bf 65}, 485 (2015),
  1602.00039.

\bibitem{Hoof:2017bqy}
GAMBIT, S.~Hoof,
\newblock {A Preview of Global Fits of Axion Models in GAMBIT},
\newblock in {\em {13th Patras Workshop on Axions, WIMPs and WISPs (AXION-WIMP
  2017) CHALKIDIKI TBC, GREECE, May 15-19, 2017}}, 2017, 1710.11138.
%%CITATION = ARXIV:1710.11138;%%

\bibitem{2013arXiv1309.7035C}
G.~{Carosi} {\em et~al.},
\newblock ArXiv e-prints  (2013), 1309.7035.

\bibitem{Raffelt:1990yz}
G.~G. Raffelt,
\newblock Phys. Rept. {\bf 198}, 1 (1990).
%%CITATION = PRPLC,198,1;%%

\bibitem{2008LNP...741...51R}
G.~G. {Raffelt},
\newblock {Astrophysical Axion Bounds},
\newblock in {\em Axions}, edited by M.~{Kuster}, G.~{Raffelt}, and
  B.~{Beltr{\'a}n}, , Lecture Notes in Physics, Berlin Springer Verlag Vol.
  741, p.~51, 2008, hep-ph/0611350.

\bibitem{2014PhRvL.113s1302A}
A.~{Ayala}, I.~{Dom{\'{\i}}nguez}, M.~{Giannotti}, A.~{Mirizzi}, and
  O.~{Straniero},
\newblock \prl {\bf 113}, 191302 (2014), 1406.6053.

\bibitem{2012aww..conf..158I}
J.~{Isern},
\newblock {White dwarfs as physics laboratories: the case of axions},
\newblock in {\em 7th Patras Workshop on Axions, WIMPs and WISPs (PATRAS
  2011)}, edited by K.~{Zioutas} and M.~{Schumann}, p. 158, 2012, 1204.3565.

\bibitem{Moody:1984ba}
J.~Moody and F.~Wilczek,
\newblock \prd {\bf 30}, 130 (1984).
%%CITATION = PHRVA,D30,130;%%

\bibitem{2012PhRvD..86a5001R}
G.~{Raffelt},
\newblock \prd {\bf 86}, 015001 (2012), 1205.1776.

\bibitem{2014PhRvL.113p1801A}
A.~{Arvanitaki} and A.~A. {Geraci},
\newblock \prl {\bf 113}, 161801 (2014), 1403.1290.

\bibitem{1983PhRvL..51.1415S}
P.~{Sikivie},
\newblock \prl {\bf 51}, 1415 (1983).

\bibitem{2010PhRvL.104d1301A}
S.~J. {Asztalos} {\em et~al.},
\newblock \prl {\bf 104}, 041301 (2010), 0910.5914.

\bibitem{2014PhRvL.112m1301S}
P.~{Sikivie}, N.~{Sullivan}, and D.~B. {Tanner},
\newblock \prl {\bf 112}, 131301 (2014), 1310.8545.

\bibitem{2016PhRvL.117n1801K}
Y.~{Kahn}, B.~R. {Safdi}, and J.~{Thaler},
\newblock Physical Review Letters {\bf 117}, 141801 (2016), 1602.01086.

\bibitem{2014PhRvX...4b1030B}
D.~{Budker}, P.~W. {Graham}, M.~{Ledbetter}, S.~{Rajendran}, and A.~O.
  {Sushkov},
\newblock Phys. Rev. X {\bf 4}, 021030 (2014), 1306.6089.

\bibitem{2017arXiv170705312G}
A.~{Garcon} {\em et~al.},
\newblock ArXiv e-prints  (2017), 1707.05312.

\bibitem{TheMADMAXWorkingGroup:2016hpc}
MADMAX Working Group, A.~Caldwell {\em et~al.},
\newblock \prl {\bf 118}, 091801 (2017), 1611.05865.
%%CITATION = ARXIV:1611.05865;%%

\bibitem{1992NuPhB.373..399E}
J.~{Ellis}, G.~B. {Gelmini}, J.~L. {Lopez}, D.~V. {Nanopoulos}, and
  S.~{Sarkar},
\newblock Nuclear Physics B {\bf 373}, 399 (1992).

\bibitem{1997PhRvD..55.7967M}
E.~{Mass{\'o}} and R.~{Toldr{\`a}},
\newblock \prd {\bf 55}, 7967 (1997), hep-ph/9702275.

\bibitem{2017JCAP...07..005C}
J.~P. {Conlon}, F.~{Day}, N.~{Jennings}, S.~{Krippendorf}, and M.~{Rummel},
\newblock \jcap {\bf 7}, 005 (2017), 1704.05256.

\bibitem{2009JCAP...08..001M}
A.~{Mirizzi}, J.~{Redondo}, and G.~{Sigl},
\newblock \jcap {\bf 8}, 1 (2009), 0905.4865.

\bibitem{2013PhRvD..88l5024T}
H.~{Tashiro}, J.~{Silk}, and D.~J.~E. {Marsh},
\newblock \prd {\bf 88}, 125024 (2013), 1308.0314.

\bibitem{2017ApJ...845L...4K}
K.~{Kelley} and P.~J. {Quinn},
\newblock \apjl {\bf 845}, L4 (2017), 1708.01399.

\bibitem{flambaum_patras}
V.~{Flambaum},
\newblock 9th Patras Workshop  (2013).

\bibitem{Abel:2017rtm}
C.~Abel {\em et~al.},
\newblock Phys. Rev. {\bf X7}, 041034 (2017), 1708.06367.
%%CITATION = ARXIV:1708.06367;%%

\bibitem{2011ConPh..52..211R}
J.~{Redondo} and A.~{Ringwald},
\newblock Contemporary Physics {\bf 52}, 211 (2011), 1011.3741.

\bibitem{2010PhLB..689..149E}
K.~{Ehret} {\em et~al.},
\newblock Physics Letters B {\bf 689}, 149 (2010), 1004.1313.

\bibitem{2013JInst...8.9001B}
R.~{B{\"a}hre} {\em et~al.},
\newblock Journal of Instrumentation {\bf 8}, 9001 (2013), 1302.5647.

\bibitem{2005PhRvL..94l1301Z}
K.~{Zioutas} {\em et~al.},
\newblock \prl {\bf 94}, 121301 (2005), hep-ex/0411033.

\bibitem{2007JCAP...04..010A}
S.~{Andriamonje} {\em et~al.},
\newblock \jcap {\bf 4}, 10 (2007), hep-ex/0702006.

\bibitem{2014PhRvL.112i1302A}
M.~{Arik} {\em et~al.},
\newblock \prl {\bf 112}, 091302 (2014), 1307.1985.

\bibitem{Vogel:2013bta}
J.~K. Vogel {\em et~al.},
\newblock {IAXO - The International Axion Observatory},
\newblock in {\em {8th Patras Workshop on Axions, WIMPs and WISPs (AXION-WIMP
  2012) Chicago, Illinois, July 18-22, 2012}}, 2013, 1302.3273.
%%CITATION = ARXIV:1302.3273;%%

\bibitem{2006PhLB..642..192A}
L.~{Amendola} and R.~{Barbieri},
\newblock Physics Letters B {\bf 642}, 192 (2006), hep-ph/0509257.

\bibitem{2015MNRAS.450..209B}
B.~{Bozek}, D.~J.~E. {Marsh}, J.~{Silk}, and R.~F.~G. {Wyse},
\newblock \mnras {\bf 450}, 209 (2015), 1409.3544.

\bibitem{2016ApJ...818...89S}
H.-Y. {Schive}, T.~{Chiueh}, T.~{Broadhurst}, and K.-W. {Huang},
\newblock \apj {\bf 818}, 89 (2016), 1508.04621.

\bibitem{2017PhRvD..95h3512C}
P.~S. {Corasaniti}, S.~{Agarwal}, D.~J.~E. {Marsh}, and S.~{Das},
\newblock \prd {\bf 95}, 083512 (2017), 1611.05892.

\bibitem{2017PhRvL.119c1302I}
V.~{Ir{\v s}i{\v c}}, M.~{Viel}, M.~G. {Haehnelt}, J.~S. {Bolton}, and G.~D.
  {Becker},
\newblock Physical Review Letters {\bf 119}, 031302 (2017), 1703.04683.

\bibitem{2017MNRAS.471.4606A}
E.~{Armengaud}, N.~{Palanque-Delabrouille}, C.~{Y{\`e}che}, D.~J.~E. {Marsh},
  and J.~{Baur},
\newblock \mnras {\bf 471}, 4606 (2017), 1703.09126.

\bibitem{bode2001}
P.~Bode, J.~P. Ostriker, and N.~Turok,
\newblock \apj {\bf 556}, 93 (2001), astro-ph/0010389.
%%CITATION = ASTRO-PH/0010389;%%

\bibitem{2016PhRvD..94l3523V}
J.~{Veltmaat} and J.~C. {Niemeyer},
\newblock \prd {\bf 94}, 123523 (2016), 1608.00802.

\bibitem{2017MNRAS.465..941D}
X.~{Du}, C.~{Behrens}, and J.~C. {Niemeyer},
\newblock \mnras {\bf 465}, 941 (2017), 1608.02575.

\bibitem{2016PhRvD..94d3513S}
B.~{Schwabe}, J.~C. {Niemeyer}, and J.~F. {Engels},
\newblock \prd {\bf 94}, 043513 (2016), 1606.05151.

\bibitem{2017MNRAS.471.4559M}
P.~{Mocz} {\em et~al.},
\newblock \mnras {\bf 471}, 4559 (2017), 1705.05845.

\bibitem{1969NCimR...1..252P}
R.~{Penrose},
\newblock Nuovo Cimento Rivista Serie {\bf 1}, 252 (1969).

\bibitem{1972Natur.238..211P}
W.~H. {Press} and S.~A. {Teukolsky},
\newblock \nat {\bf 238}, 211 (1972).

\bibitem{1973ApJ...185..649P}
W.~H. {Press} and S.~A. {Teukolsky},
\newblock \apj {\bf 185}, 649 (1973).

\bibitem{2015PhRvD..91h4011A}
A.~{Arvanitaki}, M.~{Baryakhtar}, and X.~{Huang},
\newblock \prd {\bf 91}, 084011 (2015), 1411.2263.

\bibitem{2011PhRvD..83d4026A}
A.~{Arvanitaki} and S.~{Dubovsky},
\newblock \prd {\bf 83}, 044026 (2011), 1004.3558.

\bibitem{2012PhRvL.109m1102P}
P.~{Pani}, V.~{Cardoso}, L.~{Gualtieri}, E.~{Berti}, and A.~{Ishibashi},
\newblock \prl {\bf 109}, 131102 (2012), 1209.0465.

\bibitem{2015CQGra..32m4001B}
R.~{Brito}, V.~{Cardoso}, and P.~{Pani},
\newblock Classical and Quantum Gravity {\bf 32}, 134001 (2015), 1411.0686.

\bibitem{2016PhRvL.116f1102A}
B.~P. {Abbott} {\em et~al.},
\newblock Physical Review Letters {\bf 116}, 061102 (2016), 1602.03837.

\bibitem{2017PhRvD..95d3001A}
A.~{Arvanitaki}, M.~{Baryakhtar}, S.~{Dimopoulos}, S.~{Dubovsky}, and
  R.~{Lasenby},
\newblock \prd {\bf 95}, 043001 (2017), 1604.03958.

\bibitem{2017PhRvD..96f4050B}
R.~{Brito} {\em et~al.},
\newblock \prd {\bf 96}, 064050 (2017), 1706.06311.

\bibitem{2014JCAP...02..019K}
A.~{Khmelnitsky} and V.~{Rubakov},
\newblock \jcap {\bf 2}, 019 (2014), 1309.5888.

\bibitem{2014PhRvD..90f2008P}
N.~K. {Porayko} and K.~A. {Postnov},
\newblock \prd {\bf 90}, 062008 (2014), 1408.4670.

\bibitem{2017IJMPD..2650063A}
A.~{Aoki} and J.~{Soda},
\newblock International Journal of Modern Physics D {\bf 26}, 1750063 (2017),
  1608.05933.

\bibitem{1969PhRv..187.1767R}
R.~{Ruffini} and S.~{Bonazzola},
\newblock Physical Review {\bf 187}, 1767 (1969).

\bibitem{1991PhRvL..66.1659S}
E.~{Seidel} and W.-M. {Suen},
\newblock \prl {\bf 66}, 1659 (1991).

\bibitem{1994PhRvL..72.2516S}
E.~{Seidel} and W.-M. {Suen},
\newblock \prl {\bf 72}, 2516 (1994), gr-qc/9309015.

\bibitem{Schive:2014hza}
H.-Y. {Schive} {\em et~al.},
\newblock \prl {\bf 113}, 261302 (2014), 1407.7762.

\bibitem{2015MNRAS.451.2479M}
D.~J.~E. {Marsh} and A.-R. {Pop},
\newblock \mnras {\bf 451}, 2479 (2015), 1502.03456.

\bibitem{2017MNRAS.472.1346G}
A.~X. {Gonz{\'a}lez-Morales}, D.~J.~E. {Marsh}, J.~{Pe{\~n}arrubia}, and L.~A.
  {Ure{\~n}a-L{\'o}pez},
\newblock \mnras {\bf 472}, 1346 (2017), 1609.05856.

\bibitem{2017MNRAS.468.1338C}
S.-R. {Chen}, H.-Y. {Schive}, and T.~{Chiueh},
\newblock \mnras {\bf 468}, 1338 (2017), 1606.09030.

\bibitem{2017JCAP...03..055H}
T.~{Helfer} {\em et~al.},
\newblock \jcap {\bf 3}, 055 (2017), 1609.04724.

\bibitem{2017PhRvL.118a1301L}
D.~G. {Levkov}, A.~G. {Panin}, and I.~I. {Tkachev},
\newblock Physical Review Letters {\bf 118}, 011301 (2017), 1609.03611.

\bibitem{2016PhRvD..94h3007C}
P.-H. {Chavanis},
\newblock \prd {\bf 94}, 083007 (2016), 1604.05904.

\bibitem{2016MPLA...3150090E}
J.~{Eby}, P.~{Suranyi}, and L.~C.~R. {Wijewardhana},
\newblock Modern Physics Letters A {\bf 31}, 1650090 (2016), 1512.01709.

\bibitem{2017arXiv171006268C}
P.-H. {Chavanis},
\newblock ArXiv e-prints  (2017), 1710.06268.

\bibitem{2016JHEP...12..066E}
J.~{Eby}, M.~{Leembruggen}, P.~{Suranyi}, and L.~C.~R. {Wijewardhana},
\newblock Journal of High Energy Physics {\bf 12}, 66 (2016), 1608.06911.

\bibitem{2017arXiv171004323K}
D.~F.~J. {Kimball} {\em et~al.},
\newblock ArXiv e-prints  (2017), 1710.04323.

\bibitem{2016JCAP...10..001G}
G.~F. {Giudice}, M.~{McCullough}, and A.~{Urbano},
\newblock \jcap {\bf 10}, 001 (2016), 1605.01209.

\bibitem{1990PhRvL..65.3233F}
K.~{Freese}, J.~A. {Frieman}, and A.~V. {Olinto},
\newblock \prl {\bf 65}, 3233 (1990).

\bibitem{1995PhRvL..75.2077F}
J.~A. {Frieman}, C.~T. {Hill}, A.~{Stebbins}, and I.~{Waga},
\newblock \prl {\bf 75}, 2077 (1995), astro-ph/9505060.

\bibitem{2017JCAP...01..023S}
V.~{Smer-Barreto} and A.~R. {Liddle},
\newblock \jcap {\bf 1}, 023 (2017), 1503.06100.

\bibitem{planck2015}
{Planck Collaboration} {\em et~al.},
\newblock AAP {\bf 594}, A20 (2016), 1502.02114.

\bibitem{2016arXiv161002743A}
K.~N. {Abazajian} {\em et~al.},
\newblock ArXiv e-prints  (2016), 1610.02743.

\bibitem{2008JCAP...08..003D}
S.~{Dimopoulos}, S.~{Kachru}, J.~{McGreevy}, and J.~G. {Wacker},
\newblock \jcap {\bf 8}, 3 (2008), hep-th/0507205.

\bibitem{1997PhRvL..78.1861L}
D.~H. {Lyth},
\newblock \prl {\bf 78}, 1861 (1997), hep-ph/9606387.

\bibitem{2007JHEP...06..060A}
N.~{Arkani-Hamed}, L.~{Motl}, A.~{Nicolis}, and C.~{Vafa},
\newblock \jhep {\bf 6}, 60 (2007), hep-th/0601001.

\end{thebibliography}

\end{footnotesize}

% ****************************************************************************
% END OF BIBLIOGRAPHY AREA
% ****************************************************************************

\end{document}